\PassOptionsToPackage{
  colorlinks,
  bookmarksopen,
  bookmarksnumbered,
  citecolor=teal,
  linkcolor=teal,
  urlcolor=teal
}{hyperref}
\documentclass[
  prxquantum,
  reprint,
  superscriptaddress,
  floatfix,
  longbibliography
]{revtex4-2}

\usepackage[utf8]{inputenc}
\usepackage[american]{babel}
\usepackage{booktabs}
\usepackage{amsmath,mathtools,amssymb,amsfonts}
\usepackage{amsthm}
\usepackage{braket}
\usepackage{bbold}
\usepackage{pifont}
\usepackage{MnSymbol}
\usepackage{multirow}

\usepackage{graphicx}
\usepackage{xcolor}
\usepackage{tikz}
\usepackage{tikz-network}
\usetikzlibrary{
  patterns,
  decorations.pathreplacing,
  calligraphy,
  shapes,
  arrows.meta,
  decorations.pathmorphing
}

\usepackage{enumitem}
\usepackage{comment}
\usepackage{ifthen}

\usepackage{orcidlink}

\usepackage[capitalise,noabbrev]{cleveref}

\newcommand{\titleinfo}{Higher-order Symmetric Quantum Mpemba Effect in Fragmented Systems}

\newcommand{\red}[1]{\textcolor{red}{#1}}
\definecolor{darkgreen}{RGB}{0,120,40}
\newcommand{\green}[1]{\textcolor{darkgreen}{#1}}

\begin{document}

\title{\titleinfo}

\author{Sreemayee Aditya~\orcidlink{0000-0002-0412-7944}}
\email{asreemay@uni-koeln.de}
\affiliation{Institut f\"ur Theoretische Physik, Z\"ulpicherstra\ss e 77a, 50937 K\"oln, Deutschland}

\author{Sara Murciano~\orcidlink{0000-0002-1638-5692}}
\affiliation{Universit\'e Paris-Saclay, CNRS, LPTMS, 91405 Orsay, France}

\author{Xhek Turkeshi~\orcidlink{0000-0003-1093-3771}}
\affiliation{Institut f\"ur Theoretische Physik, Z\"ulpicherstra\ss e 77a, 50937 K\"oln, Deutschland}

\begin{abstract}
A quantum system can restore a broken symmetry faster the more strongly it initially breaks it, an anomaly known as the quantum Mpemba effect. Whether this effect survives once conservation laws fragment the Hilbert space into exponentially many disconnected Krylov sectors has remained open. We address this question for circuits and Hamiltonians with simultaneous charge and dipole conservation, the paradigmatic setting for strong Hilbert-space fragmentation. Combining a replica tensor-network formulation for charge and dipole-conserving gates, which reaches the annealed R\'enyi-2 entanglement asymmetry up to $L=128$, with Hamiltonian simulations and an exactly solvable dissipative model, we uncover a higher-order symmetric quantum Mpemba effect: the charge and dipole asymmetries each display Mpemba-like crossings on parametrically distinct timescales. Resolving the state into frozen and active Krylov sectors reveals the mechanism: frozen fragments retain a finite asymmetry that obstructs full restoration, while active fragments host the relaxation responsible for the crossings. Fragmentation thus does not preclude the quantum Mpemba effect but reshapes it into frozen memory and active-fragment relaxation, providing a framework for the Mpemba phenomenology of higher-moment symmetries.
\end{abstract}

\maketitle

\section{Introduction}
The Mpemba effect refers to the counter-intuitive phenomenon in which a system prepared farther from equilibrium relaxes faster than one prepared closer to it~\cite{mpemba1969cool}.
Originally observed in classical heat transport~\cite{lasanta2017hotter,lu2017nonequilibrium,klich2019mpemba,kumar2020exponentially,bechhoefer2021fresh,kumar2022anomalous}, it is now understood as a broad feature of nonequilibrium relaxation~\cite{polkovnikov_2011,rigol2008thermalization,D_Alessio_2016}.
Its quantum analogue, the quantum Mpemba effect (QME), was first studied in open quantum systems~\cite{carollo2021exponentially}, and has since been recognized as a far more general phenomenon.
A unifying perspective has recently emerged from quantum resource theories~\cite{summer2025resourcetheoreticalunificationmpemba}.
Just as the original thermodynamic effect concerns the distance from equilibrium relaxing toward the Gibbs ensemble, the QME can be cast more generally as the anomalous relaxation of a resource monotone: a state initially richer in a given quantum resource flows toward the resource-free manifold faster than a state initially poorer in it.
Different choices of resources and free operations recover different incarnations of the effect, including abstract quantum information metrics such as nonstabilizerness or coherences~\cite{Aditya25Mbempa,xiao2026nonstabilizernessmpembaeffects,Kusuki2026end,Travaglino2026}.

Within the framework of closed many-body systems, the most studied form is the \emph{symmetric} quantum Mpemba effect (SQME), where the relevant resource is \textit{asymmetry}, namely the breaking of a global symmetry.
This is diagnosed through the anomalous symmetry restoration~\cite{ares2023entanglement,ares2023lack,yamashika2024quantum,Chalas2024,Bertini2024,turkeshi2025quantum,liu2024symmetry,Ares2025}: an initial state that breaks a symmetry more strongly can restore it faster than a less asymmetric one.
The natural monotone here is the relative entropy of asymmetry, or entanglement asymmetry for short~\cite{ares2023entanglement,vaccaro2008tradeoff,gour2009measuring}, a non-negative functional of the reduced density matrix that quantifies how far a state is from the symmetric manifold and vanishes if and only if the symmetry is restored.
The symmetric QME have been investigated in integrable models~\cite{rylands-24,ares2023entanglement,amvc-23,Chalas2024,Murciano2024EntanglementAsymmetry,yamashika2024quantum,Russotto_2026,rylands2024,Yamashika24}, conformal field theories~\cite{Benini2025EntanglementAsymmetryHolography}, dissipative and monitored systems~\cite{Ares2025,caceffo2024entangled,DiGiulio2025measurement,Zeng2026ylj}, random circuits~\cite{turkeshi2025quantum,liu2024symmetry,summer2025resourcetheoreticalunificationmpemba,Li26longrange,foligno25}, long-range spin chains~\cite{yamashika25longrange,Yu2025TuningQuantumMpemba,Hallam2025}, quantum simulators~\cite{joshi-24}, and chaotic systems~\cite{bhore25chaotic,muller2026quantummpembaeffectchaotic}; see Refs.~\cite{Ares2025QuantumMpembaEffects,teza2025speedupsnonequilibriumthermalrelaxation,Calabrese2026} for a review of the quantum and classical aspects of this phenomenon.

However, most studies of the SQME so far have focused on ordinary $U(1)$ charge conservation, leaving open how the effect behaves under richer symmetry structures.
A natural arena is provided by constrained quantum systems~\cite{Moudgalya_review_2022}, which obey higher-rank conservation laws where not only the total charge but also its moments, such as the dipole moment, are conserved~\cite{pretko17highermoments}.
Such constraints generically lead to Hilbert-space fragmentation (HSF)~\cite{Moudgalya_pairhopping_2020,sala_dipole,sala_ergo_2020,khemani_2020,Moudgalya_review_2022,aditya_2023,Deepak_HSF,East_Sreemayee,Aditya25higherEast,mukherjee_2021}: even a fixed charge-and-dipole sector decomposes into exponentially many dynamically disconnected Krylov fragments~\cite{Moudgalya_review_2022}.
These fragments include an exponentially large number of one-dimensional frozen sectors, consisting of configurations that do not evolve under the dynamics, as well as active sectors, in which multiple configurations are connected by the allowed local moves. In the strong-HSF regime, this fragmentation is even more severe: the largest fragment contained within a conventional symmetry sector remains exponentially small in the thermodynamic limit~\cite{sala_dipole,Moudgalya_review_2022,khemani_2020}. As a consequence, fragmentation can strongly affect the dynamics following a quantum quench. 
Frozen fragments, being dynamically inactive, retain memory of the initial state, whereas active fragments relax extremely slowly; different fragments can moreover display very different relaxation times and even qualitatively distinct late-time behavior, giving rise to dynamical heterogeneity~\cite{Lenart_hetero,Causer24Fredkin,maric26slwodynamics}.
Each of these features bears directly on symmetry restoration, making it natural to ask how strong HSF reshapes the QME when several global symmetries, here charge and dipole conservations, must be restored at once.

\begin{figure}[t]
\centering

\includegraphics[width=\columnwidth]{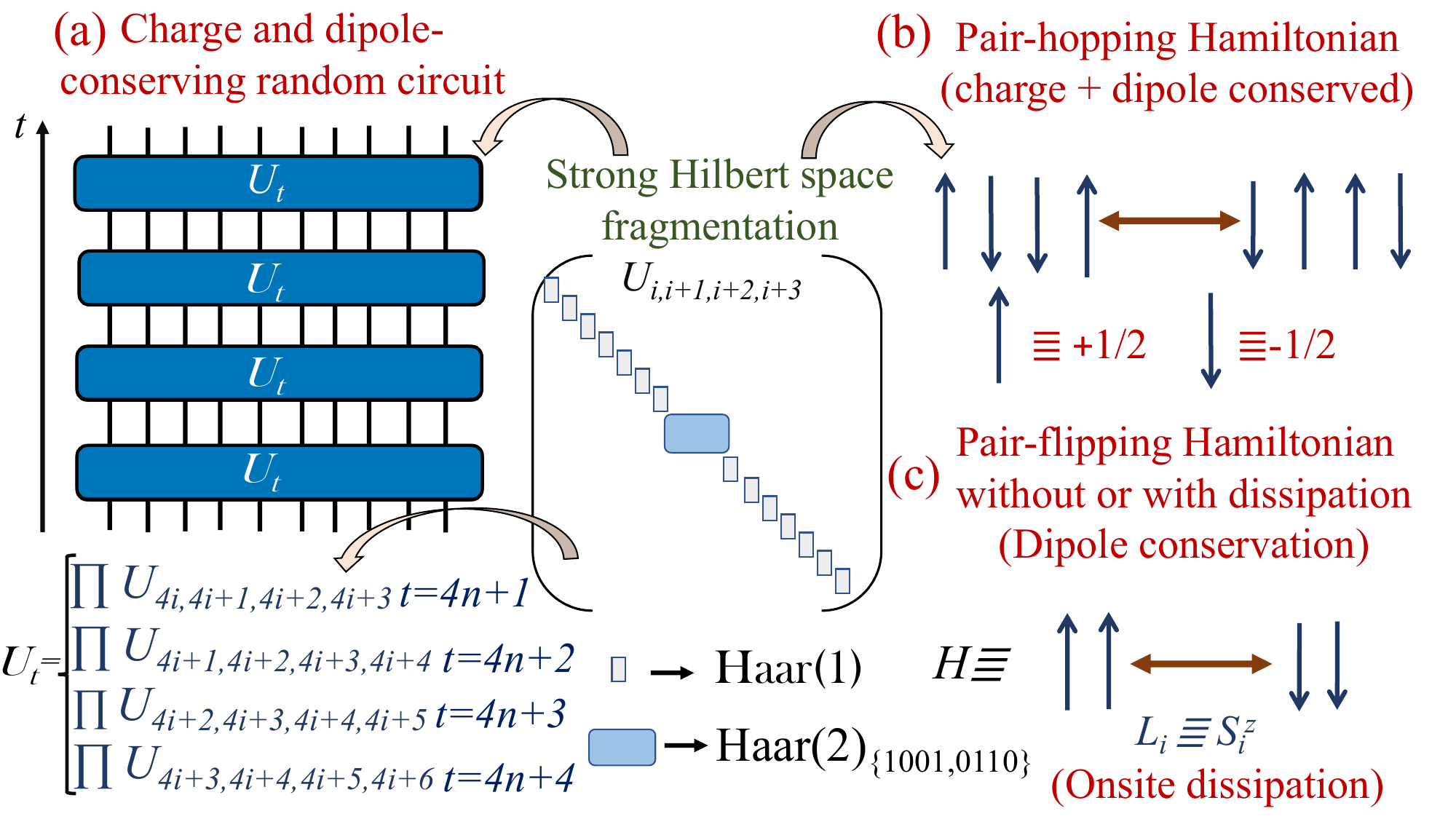}

\vspace{1mm}

\renewcommand{\arraystretch}{1.10}
\setlength{\tabcolsep}{2.0pt}
\scriptsize
\begin{tabular}{|p{0.24\columnwidth}|p{0.29\columnwidth}|p{0.41\columnwidth}|}
\hline
~~~~~~~\textbf{Setup} & ~~~~~~~~\textbf{Method} & ~~~~~~~~~~~~~~\textbf{Results} \\
\hline

Random circuit
&
RTN for $L\leq128$
&
QME present, both for tilted ferromagnetic and antiferromagnetic states; strong HSF yields crossings in both sectors and an asymptotically slow approach to symmetry restoration.
\\
\hline

Pair hopping
&
Exact vector simulation; Chebyshev propagation for $L\leq20$
&
QME present, with the same underlying mechanism as in the random-circuit case.
\\
\hline

Pair-flipping model with dissipation $\gamma$
&
Factorized Lindblad dynamics; closed-form expression for charge and dipole asymmetry
&
For $\gamma>0$, multiple crossings with symmetry restoration in both sectors; the smallest crossing time is
$t_M=\arcsin[1/\sqrt{\sin^2\theta_1+\sin^2\theta_2}]$.
\\
\hline

\end{tabular}

\vspace{-1mm}

\caption{Schematic and summary of our study.}
\label{fig:schematic_with_summary}

\end{figure}

In this work, we address these questions within archetypal frameworks that combines extensive numerical simulations with analytical methods, employing the entanglement asymmetry as a proxy for the Mpemba phenomenology.
Across the three settings considered below, we uncover a \textit{higher-order SQME}, in which the entanglement asymmetries associated with charge and dipole conservation each display Mpemba-like crossings, with parametrically distinct relaxation scales.
To disentangle universal features from model-specific artifacts, we first consider the dynamics of random quantum circuits on spin-$1/2$ degrees of freedom that preserve both the total charge and the dipole moment~\cite{khemani_2020}.
These minimal models retain only the structure imposed by unitarity, locality, and symmetry, and therefore expose the generic phenomenology of fragmented systems. A central numerical challenge is that the asymptotic regime relevant for the asymmetry emerges only at large system sizes, whereas exact vector simulation is confined to chains of a handful of spins, well below the scale at which the asymptotic fragmentation signature fully sets in. To overcome this bottleneck, rather than tackling individual circuit realizations we consider the ensemble-averaged dynamics of the entanglement asymmetry. This average maps onto a classical statistical-mechanics model that we resum exactly through a replica tensor-network (RTN) formulation adapted to charge- and dipole-conserving gates, extending earlier constructions for charge-conserving circuits~\cite{curt,khemani,Rakovszky2018DiffusiveHydrodynamics,turkeshi2025quantum,aditya2026coherencedynamicsquantummanybody}. This formalism grants access to the annealed entanglement asymmetry for systems as large as $L=128$, well beyond the reach of exact vector simulation. Within the accessible time window, symmetry restoration in the random circuit appears either incomplete or asymptotically slow, both for subsystems attached to the boundary and for those embedded in the bulk. To isolate the role of fragmentation, we resolve
the random-circuit dynamics within each Krylov sector. This decomposition reveals a transparent mechanism: frozen fragments retain a finite asymmetry that does not decay, while active fragments host the relaxation processes responsible for the Mpemba-like crossings. Hilbert-space fragmentation therefore does not preclude the QME but reshapes it, splitting symmetry restoration into a frozen-memory contribution and an active-fragment relaxation channel. Because this frozen-plus-active decomposition relies only on the existence of dynamically disconnected Krylov sectors (and not on the specific conservation laws involved) the same mechanism applies to any constrained model with a comparable fragmentation structure. The main findings of our analysis are summarized in Fig.~\ref{fig:schematic_with_summary}, and provide a systematic framework for the QME associated with higher-moment symmetries.

We next consider the dynamics generated by a charge- and dipole-conserving pair-hopping Hamiltonian~\cite{Moudgalya_pairhopping_2020}.
Energy conservation introduces a physically motivated additional structure that may reshape both the relaxation pathways and the relevant timescales, as recently shown in the context of general quantum resource theories~\cite{aditya2025growthspreadingquantumresources,aditya2026coherencedynamicsquantummanybody,tirrito25anticoncentration,turkeshi2025magic}; whether the higher-order symmetric Mpemba effect persists under energy-conserving evolution is therefore a question of physical relevance. In this setting circuit averaging is no longer available to recast the dynamics as a replica tensor network. We instead adapt the state-of-the-art Chebyshev polynomial method~\cite{TalEzer1984AccurateEfficientScheme,Sierant2022ChallengesMBL} to the Hamiltonian evolution, which again enables the largest-scale available numerics via vector simulation methods.

While the numerical analysis already establishes a compelling phenomenology, we complement it with a minimal toy model that captures the key features and admits an exact solution.
The model combines the minimal symmetric pair-flipping interaction with local on-site dephasing. The dynamics is now explicitly dissipative, but the dephasing captures, within a Markovian approximation, the effective action of the complementary subsystem on the entanglement asymmetry, the role of the ``bath'' associated with the bipartition.
This bath effect is essential: in the absence of local dephasing the symmetry is not restored and no SQME emerges. The model yields closed-form expressions for the asymmetries that corroborate the phenomenology established numerically.

\section{Setup}

In this section we set up our study. We first introduce the three settings in which we probe the higher-order QME (a charge- and dipole-conserving random circuit, the corresponding pair-hopping Hamiltonian, and an analytically tractable dissipative pair-flip model) and then define the entanglement asymmetry and the quantum Mpemba criterion, together with the numerical methods tailored to each setting.

\subsection{The models}

\textit{Charge- and dipole-conserving circuit.} Our first setting is a chain of $L$ spin-$1/2$ degrees of freedom, $\mathcal{H}=(\mathbb{C}^{2})^{\otimes L}$, evolving under a random brickwork circuit that conserves both the total $U(1)$ charge $Q=\sum_{x} S^z_x$ and the dipole moment $P=\sum_{x} x\, S^z_x$, with $S^z_x=\sigma^z_x/2$. Dipole conservation is an inhomogeneous symmetry: beyond fixing the number of up and down spins it also constrains their positions, and is the microscopic origin of the strong Hilbert-space fragmentation discussed above. The constraint is so severe that no nontrivial move fits within two or three sites, where any two configurations of equal charge necessarily differ in dipole moment. The minimal nontrivial gate therefore spans four sites, on which the only allowed rearrangement is the pair hopping
\begin{equation}
\ket{+1/2,-1/2,-1/2,+1/2}
\leftrightarrow
\ket{-1/2,+1/2,+1/2,-1/2},
\label{eq:pairhop}
\end{equation}
between two configurations of identical charge $Q$ and dipole moment $P$. A charge- and dipole-conserving four-site gate is thus block diagonal: it acts as a Haar-random $U(2)$ rotation on this two-dimensional subspace, while the remaining fourteen basis states form frozen one-dimensional blocks carrying independent random phases $\phi_\alpha$,
\begin{eqnarray}
  U_{i:i+3}
  &=& \bigoplus_{\alpha}
    e^{i\phi_\alpha}\ket{\alpha}\bra{\alpha}
  \;\nonumber\\&&\oplus\;
  \begin{pmatrix}
    u_{11} & u_{12}\\[2pt]
    u_{21} & u_{22}
  \end{pmatrix}_{\!\{\ket{-1/2,+1/2,+1/2,-1/2},\,\ket{+1/2,-1/2,-1/2,+1/2}\}},
  \label{eq:gate}
\end{eqnarray}
with $\alpha$ labelling the fourteen frozen states. Time evolution proceeds by a brickwork circuit of four staggered layers per Floquet step, $\mathcal{U} = \mathcal{U}^{(4)}\mathcal{U}^{(3)}\mathcal{U}^{(2)}\mathcal{U}^{(1)}$, with
\begin{align}\label{eq:layers}
  \mathcal{U}^{(a)} &= \prod_{j\equiv a\;\mathrm{mod}\;4} U_{j,j+1,j+2,j+3}\,, \qquad a=1,2,3,4.
\end{align}
Each gate is drawn independently from the charge- and dipole-conserving unitary group, under open boundary conditions (OBCs), so that after $t$ steps the state reads $\rho(t) = \mathcal{U}^t\,\rho_0\,(\mathcal{U}^\dagger)^t$.

\textit{Charge- and dipole-conserving Hamiltonian.} 
As a second setting, we consider a coherent continuous-time evolution under a Hamiltonian with the same two conservation laws. Built from the elementary pair-hopping process of Eq.~\eqref{eq:pairhop}, the minimal such Hamiltonian reads
\begin{equation}
  H = J\sum_{i=1}^{L-3}\big(
    S^+_i S^-_{i+1} S^-_{i+2} S^+_{i+3}
    + \mathrm{h.c.}\big)\,,
\label{eq:ham_dipole}
\end{equation}
with $S^\pm_i = (X_i \pm i Y_i)/2$ the spin raising and lowering operators and $J$ the pair-hopping scale. By construction $[H,Q]=[H,P]=0$; we again employ OBCs, and the state evolves as $\rho(t) = e^{-iHt}\,\rho_0\,e^{iHt}$.

\textit{Symmetric pair-flip model.} To make analytical progress, we introduce a toy model that retains the essential structure of dipole-conserving dynamics. Measuring the dipole relative to the chain center $j_0 = L/2$, $P = \sum_{x}(x - j_0)\, S^z_x$, the Hamiltonian flips pairs of spins placed symmetrically about $j_0$,
\begin{equation}\label{eq:ham_sym}
  H = J\sum_{x=1}^{L/2}\big(
    S^+_{j_0+x}\,S^+_{j_0-x} + \mathrm{h.c.}
  \big)\,.
\end{equation}
The $+x$ and $-x$ contributions to the dipole cancel, so $P$ is exactly conserved, $[H,P]=0$, whereas the total charge is not: each term shifts $Q$ by $\pm2$, leaving only the parity $(-1)^Q$ invariant. On its own this model is too simple to thermalize, producing mere coherent oscillations between paired configurations; we therefore couple it to a local dephasing bath, so that the density matrix obeys the Gorini--Kossakowski--Sudarshan--Lindblad (GKSL) master equation~\cite{BreuerPetruccione2002OpenQuantumSystems}
\begin{equation}\label{eq:lindblad}
  \frac{d\rho}{dt} = -i[H,\rho]
  + \gamma\sum_{x=1}^{L}\bigg(
    L_x\,\rho\,L_x^\dagger
    - \frac{1}{2}\big\{L_x^\dagger L_x,\,\rho\big\}
  \bigg)\,,
\end{equation}
with jump operators $L_x = S^z_x$ and dephasing rate $\gamma\geq 0$. Since $[S^z_x, Q]=[S^z_x, P]=0$, the dissipation preserves the full symmetry structure of the coherent dynamics.
 
\subsection{Entanglement asymmetry and quantum Mpemba criterion}
To diagnose the QME in the presence of several conserved quantities, we track the relaxation of the entanglement asymmetry~\cite{ares2023entanglement}, which quantifies how strongly a state breaks a given symmetry, applied separately to the global charge and the dipole.
Consider a bipartition of the chain into a subsystem $A$ and its
complement $B$. Given the reduced density matrix $\rho_A(t)=\mathrm{tr}_B[\rho(t)]$, we define the symmetrized state with respect to the subsystem operator $\mathcal O_A$ as
\begin{equation}
\rho_{A,\mathcal O}(t)
=
\sum_o \Pi^{\mathcal O_A}_o\,\rho_A(t)\,\Pi^{\mathcal O_A}_o .
\end{equation}
Here $\Pi^{\mathcal O_A}_o$ is the projector onto the eigenspace of
$\mathcal O_A$ with eigenvalue $o$. 
In what follows we will consider two choices:
the local charge
\begin{equation}
Q_A=\sum_{x\in A} S_x^z ,
\end{equation}
and the local dipole moment
\begin{equation}
P_A=\sum_{x\in A} x\,S_x^z .
\end{equation}
For any R\'enyi index $n$, we define the entanglement asymmetry
associated with $\mathcal O=Q,P$ as
\begin{equation}
\label{eq:asymmetry}
\Delta S^{(n)}_{\mathcal O}(\rho_A)
=
S_n(\rho_{A,\mathcal O})-S_n(\rho_A),
\end{equation}
where
\begin{equation}
S_n(\rho)
=
\frac{1}{1-n}\ln \mathrm{tr}(\rho^n).
\end{equation}
In the limit $n\to 1$, this reduces to the von Neumann entanglement
asymmetry~\cite{HolzheyLarsenWilczek1994GeometricEntropy,CalabreseCardy2004EntanglementEntropy},
\begin{equation}
\Delta S^{v}_{\mathcal O}(\rho_A)
=
S^{v}(\rho_{A,\mathcal O})-S^{v}(\rho_A),
\qquad
S^{v}(\rho)=-\mathrm{tr}(\rho\ln\rho).
\end{equation}
The quantity $\Delta S^{(n)}_{\mathcal O}$ is non-negative and
vanishes whenever the symmetry is restored. 

We diagnose the quantum Mpemba effect directly through the R\'enyi-$n$ entanglement asymmetry $\Delta S^{(n)}_{\mathcal O}$\footnote{A word of caution on monotonicity is in order. The genuine resource monotone for asymmetry is the von Neumann (relative-entropy) measure, which is non-increasing under \emph{every} charge-conserving (covariant) operation, including measurements averaged over their outcomes. The R\'enyi-$n$ surrogates with $n>1$ that we employ (chosen because they require only integer moments of $\rho_A$ and are therefore analytically, numerically, and experimentally accessible~\cite{joshi-24}) are invariant under covariant \emph{unitary} dynamics, but need not be monotone under irreversible covariant maps: Ref.~\cite{Kusuki2026end} exhibits a covariant (dephasing/measurement-type) channel under which $\Delta S^{(2)}$ strictly increases. In other words, the surrogate is well behaved precisely when one does not demand monotonicity under such on-average measurement operations. Since the quenches studied here are generated by covariant unitary dynamics (the reduced-state asymmetry changing only through the partial trace) $\Delta S^{(n)}$ remains a faithful Mpemba diagnostic, which we cross-check at $n\to1$. A complete resource-theoretic characterization of the R\'enyi asymmetries, in particular under measurement-type operations, is left open.}. Given two initial states $\rho(0)$ and $\sigma(0)$, we say that $\rho(0)$ is initially more asymmetric than $\sigma(0)$ with respect to $\mathcal O$ if
\begin{equation}
\Delta S^{(n)}_{\mathcal O}(\rho_A(0))
>
\Delta S^{(n)}_{\mathcal O}(\sigma_A(0)).
\end{equation}
The QME for the operator $\mathcal O$ occurs when this initial ordering is reversed during the relaxation, i.e.\ when there exists a Mpemba time $t_{M}$~\cite{rylands-24} such that
\begin{align}
\Delta S^{(n)}_{\mathcal O}(\rho_A(0))
&>
\Delta S^{(n)}_{\mathcal O}(\sigma_A(0)),
\label{eq:mpemba_criterion}\\
\Delta S^{(n)}_{\mathcal O}(\rho_A(\tau))
&<
\Delta S^{(n)}_{\mathcal O}(\sigma_A(\tau)),
\qquad
\tau>t_{M}.
\end{align}
The criterion applies independently to the charge and dipole asymmetries, so that the restoration of the two conservation laws can display Mpemba-like behavior either simultaneously or separately.

\subsection{Asymmetry of the initial state}
To build intuition for the crossing criterion of Eq.~\eqref{eq:mpemba_criterion}, it is instructive to first quantify the asymmetry carried by the initial states themselves. Throughout the main text we take the tilted ferromagnetic product state~\cite{ares2023entanglement,ares2023lack,Ares2025QuantumMpembaEffects}
\begin{equation}
\ket{\Psi^{\mathrm f}_0(\theta)}
=
\bigotimes_{i=1}^{L}
e^{-iY_i\theta/2}\ket{+1/2},
\label{eq:tilted_ferro}
\end{equation}
i.e.\ each spin carries amplitude $\cos(\theta/2)$ on $\ket{+1/2}$ and $\sin(\theta/2)$ on $\ket{-1/2}$. This is the prototypical symmetry-breaking initial condition, controlled by a single tilt angle $\theta$; other choices, such as the tilted antiferromagnet, yield the same picture and are treated in Appendix~\ref{app:tilted_AFM}.

For such a product state the asymmetry can be computed in closed form. The reduced density matrix on the boundary subsystem $A=\{1,\dots,L_A\}$ is pure, so every symmetry-projected block stays pure whenever its weight is nonzero, and the asymmetry collapses onto the Shannon (or, for $n=2$, the R\'enyi-$2$) entropy of the distribution $p_q$ of the conserved charge over $A$,
\begin{equation}
\Delta S_{\mathcal O}=-\sum_{q} p_q\log p_q,
\qquad
\Delta S^{(2)}_{\mathcal O}=-\log\!\Big(\sum_{q} p_q^2\Big).
\label{eq:initial_asymmetry_product}
\end{equation}
The sector weights $p_q$ follow from the character representation of the projector onto the eigenspace of the conserved operator $\mathcal O_A$,
\begin{equation}
p_q
=
\int_{-\pi}^{\pi}\frac{d\alpha}{2\pi}\,
\bra{\Psi^{\mathrm f}_0(\theta)}\,e^{i\alpha \mathcal O_A}\,\ket{\Psi^{\mathrm f}_0(\theta)}\,
e^{-i\alpha q}.
\label{eq:initial_sector_weight}
\end{equation}
Setting $a=\cos^2(\theta/2)$ and $b=\sin^2(\theta/2)$, the charge $Q_A=\sum_{x\in A}S^z_x$ factorizes over sites into a binomial distribution,
\begin{equation}
p^{(Q)}_q
=
\binom{L_A}{n}\,a^{n}\,b^{L_A-n},
\qquad
n=q+\tfrac{L_A}{2},
\label{eq:initial_charge_weight}
\end{equation}
whereas the dipole $P_A=\sum_{x\in A}x\,S^z_x$ produces a site-weighted generating function,
\begin{equation}
p^{(P)}_q
=
\int_{-\pi}^{\pi}\frac{d\alpha}{2\pi}
\prod_{x=1}^{L_A}\big(a\,e^{i\alpha x/2}+b\,e^{-i\alpha x/2}\big)\,
e^{-i\alpha q}.
\label{eq:initial_dipole_weight}
\end{equation}
Inserting these weights into Eq.~\eqref{eq:initial_asymmetry_product} yields the charge and dipole asymmetries in closed form.

\begin{figure*}
\includegraphics[width=0.98\textwidth]{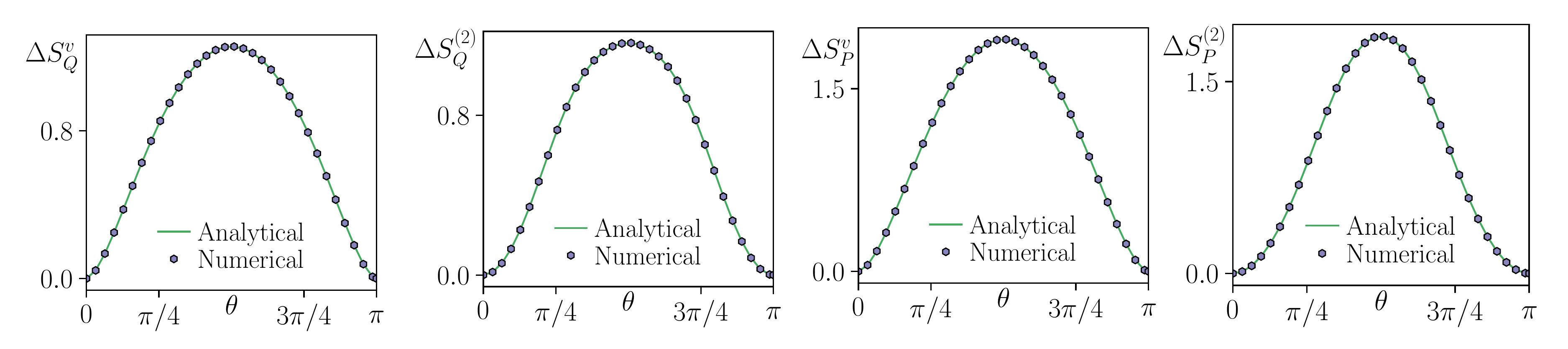}
\caption{
Von Neumann and R\'enyi-2 asymmetries associated with charge and dipole  for an initial pure spin-$1/2$ tilted ferromagnetic state, computed for a boundary subsystem of size $L_A=3$ as a function of the tilting angle.
In all cases, the analytical results are in excellent agreement with the numerical data.
}
\label{fig:initial_asymmetry}
\end{figure*}

As shown in Fig.~\ref{fig:initial_asymmetry}, both grow monotonically with the tilt angle on $\theta\in[0,\pi/2]$, in excellent agreement with the numerics: a more strongly tilted state breaks the symmetry more and therefore starts farther from the symmetric manifold. This monotonicity is what makes the QME simple to set up. To enforce the initial ordering required by Eq.~\eqref{eq:mpemba_criterion}, it suffices to prepare two states at two different angles $\theta_1\neq\theta_2$, the more tilted one being the more asymmetric; whether the relaxation then reverses this ordering (the defining signature of the QME) is the question we address next.

\subsection{Numerical methods}
The three settings require distinct computational strategies. We organize them around two complementary tools: \emph{exact vector simulation}, used for the two Hamiltonian settings and as a benchmark, and a \emph{replica tensor network}, which delivers the circuit-averaged asymmetries of the random circuit at large $L$.

\textit{Exact vector simulation.} For the charge- and dipole-conserving Hamiltonian [Eq.~\eqref{eq:ham_dipole}] we evolve the state with a Chebyshev polynomial expansion of $e^{-iHt}$~\cite{TalEzer1984AccurateEfficientScheme,Sierant2022ChallengesMBL}, which is numerically exact and reaches $L\le20$. For the symmetric pair-flip model [Eq.~\eqref{eq:ham_sym}] each term couples only a mirror pair of sites, so the evolution factorizes into independent two-site blocks; the same factorization carries over to the dissipative Lindblad dynamics [Eq.~\eqref{eq:lindblad}], because the dephasing jumps $S^z_x$ do not mix sectors, and it is what makes the closed-form expressions for the dissipative model possible. The random circuit can likewise be evolved exactly by sampling Haar gates realization by realization; this is restricted to a handful of spins but lets us establish the self-averaging of the asymmetry, i.e.\ that the annealed average tracks the typical (quenched) value (Appendix~\ref{app:ED_RTN_benchmark}).

\textit{Replica tensor network.} For the charge- and dipole-conserving random circuit, exact vector simulation reaches only small chains, so we instead target the \emph{annealed} averages directly with a replica tensor network~\cite{Rakovszky2018DiffusiveHydrodynamics,turkeshi2025quantum,aditya2026coherencedynamicsquantummanybody}, see Ref.~\cite{turkeshi2026lecturenotesreplicatensor} for a pedagogical introduction.
The central objects are the circuit-averaged purities
$\mathcal{P}^{(n)}_{A}
= \mathbb{E}[\mathrm{tr}(\rho_A^n)]$,
$\mathcal{P}^{(n)}_{A,Q}
= \mathbb{E}[\mathrm{tr}(\rho_{A,Q}^n)]$, and
$\mathcal{P}^{(n)}_{A,P}
= \mathbb{E}[\mathrm{tr}(\rho_{A,P}^n)]$,
from which the annealed charge and dipole
entanglement asymmetries are obtained as
\begin{align}\label{eq:annealed_QP}
  \widetilde{\Delta S}^{(n)}_{\mathcal{O}}
  &= \frac{1}{1-n}\ln\!\bigg[
    \frac{\mathcal{P}^{(n)}_{A,\mathcal{O}}}
         {\mathcal{P}^{(n)}_{A}}
  \bigg]\,,\qquad \mathcal{O}=Q,P.
\end{align}
Since each gate is independently drawn, the Haar
average can be performed gate by gate. For $n=2$,
this yields a local transfer operator
$\mathcal{T}_{i:i+3}
= \mathbb{E}[(U_{i:i+3}\otimes U^*_{i:i+3})^{\otimes 2}]$
via the Haar fourth-moment identity. Although related tensor-network methods exist for charge-conserving circuits, the extension to the four-site dipole-conserving gate in Eq.~\eqref{eq:gate} is technically nontrivial. The dipole constraint induces a highly fragmented local structure, making large-$L$ simulations challenging; the formulation developed here overcomes this obstacle and gives access to the asymptotic strong-HSF regime. Even though the detailed construction is given in
Appendix~\ref{app:rtn_four_site}, we summarize the main ideas here.

\begin{figure}[t]
\centering

\definecolor{rtnT}{RGB}{31,119,180}
\definecolor{rtnG}{RGB}{29,158,117}
\definecolor{rtnS}{RGB}{150,150,140}
\definecolor{rtnW}{RGB}{216,90,48}
\definecolor{rtnL}{RGB}{120,120,120}

\resizebox{\columnwidth}{!}{%
\begin{tikzpicture}[
  every node/.style={font=\large},
  leg/.style={rtnL,line width=0.70pt},
  tbox/.style={
    draw=rtnT!60!black,
    fill=rtnT!12,
    text=black,
    rounded corners=1.5pt,
    minimum height=5.5mm,
    font=\Large,
    inner sep=1.8pt
  },
  state/.style={
    draw=rtnS!70!black,
    fill=rtnS!18,
    text=black,
    rounded corners=1.5pt,
    minimum height=5.5mm,
    font=\Large,
    inner sep=1.8pt
  },
  swap/.style={
    draw=rtnW!60!black,
    fill=rtnW!14,
    text=black,
    rounded corners=1.5pt,
    minimum height=5.5mm,
    font=\Large,
    inner sep=1.8pt
  },
  gbox/.style={
    draw=rtnG!55!black,
    fill=rtnG!12,
    text=black,
    rounded corners=1.5pt,
    minimum height=5.5mm,
    font=\Large,
    inner sep=1.8pt
  }]

\def\ybot{0.0}
\def\ytop{5.55}

\draw[-{Stealth[length=2.2mm]},thick] (-0.55,0.0) -- (-0.55,5.75)
  node[left,pos=0.99] {$t$};

\foreach \x in {1,...,12}{
  \draw[leg] (\x,\ybot) -- (\x,\ytop);
}

\node[state,minimum width=11.6cm] at (6.5,-0.36)
  {$\lvert\rho_0^{\otimes 2}\rangle\!\rangle$};

\foreach \x in {1,...,12}{
  \draw[leg] (\x,-0.10) -- (\x,0.0);
}

\newcommand{\Tfullsmall}[2]{%
  \node[tbox,minimum width=3.40cm] at ({#1+1.5},#2)
  {$\mathcal{T}$};}

\newcommand{\tstepsmall}[2]{%
  \node[anchor=center] at (0.12,#2) {$t=#1$};}

\tstepsmall{1}{0.82}
\Tfullsmall{1}{0.82}
\Tfullsmall{5}{0.82}
\Tfullsmall{9}{0.82}

\tstepsmall{2}{1.78}
\Tfullsmall{2}{1.78}
\Tfullsmall{6}{1.78}

\tstepsmall{3}{2.74}
\Tfullsmall{3}{2.74}
\Tfullsmall{7}{2.74}

\tstepsmall{4}{3.70}
\Tfullsmall{4}{3.70}
\Tfullsmall{8}{3.70}

\tstepsmall{5}{4.45}
\node at (6.5,4.45) {$\vdots$};

\node[
  anchor=west,
  draw=rtnS!70!black,
  fill=rtnS!10,
  rounded corners=1pt,
  inner sep=1.2pt
] at (-0.3,-1.1)
  {$i=t\bmod 4$,OBCs};

\foreach \x in {1,...,12}{
  \draw[leg] (\x,5.05) -- (\x,5.55);
}

\node[swap,minimum width=3.15cm] at (2.0,5.35)
  {$F_A$};

\node[state,minimum width=8.05cm] at (8.0,5.35)
  {$\mathbb{1}_{\bar A}$};

\draw[decorate,decoration={brace,amplitude=2.5pt,mirror}]
  (0.78,4.82) -- (3.22,4.82);

\node[align=center,anchor=north] at (2.0,4.75)
  {$A$, $L_A=3$};

\node[anchor=north,align=center] at (7.8,-0.96)
  {$\displaystyle
  \mathcal{P}^{(2)}_{A,\mathcal{O}}(t)=
  \langle\!\langle F^{(\mathcal{O})}_{A}\rvert\,
  \mathcal{T}^{(t\bmod4)}\cdots\mathcal{T}^{(1)}
  \lvert\rho_0^{\otimes2}\rangle\!\rangle$};

\begin{scope}[shift={(0.35,-3.80)}]

\foreach \x in {1,...,4}{
  \draw[leg] (\x*0.50,0.35) -- (\x*0.50,1.75);
}

\node[
  gbox,
  minimum width=3.05cm,
  minimum height=6.0mm
] at (1.25,1.05)
  {$U_{i,\dots,i+3}$};

\draw[-{Stealth[length=2.4mm]},thick] (3.05,1.05) -- (4.35,1.05)
  node[midway,above]
  {$\mathbb{E}_{\mathrm{Haar}}$};

\foreach \x in {1,...,4}{
  \draw[leg] (\x*0.50+4.55,0.35) -- (\x*0.50+4.55,1.75);
}

\node[
  tbox,
  minimum width=3.05cm,
  minimum height=6.0mm
] at (5.80,1.05)
  {$\mathcal{T}_{i,\dots,i+3}$};

\node[anchor=west,align=left] at (7.55,1.05)
  {$\mathcal{O}$ = swap with no twist: full purity\\
   $\mathcal{O}$ = charge-twisted swap: $Q$\\
   $\mathcal{O}$ = dipole-twisted swap: $P$};

\node[
  anchor=north,
  align=center,
  font=\Large,
  text width=11.5cm
] at (5.0,-0.05)
  {$\displaystyle
  \mathcal{T}_{i,\dots,i+3}
  =
  \sum_{\alpha_1,\alpha_2}
  \left[\frac{
  \lvert I_{\alpha_1\alpha_2}\rangle\!\rangle
  \langle\!\langle I_{\alpha_1\alpha_2}\rvert
  }{d_{\alpha_1}d_{\alpha_2}}
  +
  \frac{
  \lvert J_{\alpha_1\alpha_2}\rangle\!\rangle
  \langle\!\langle J_{\alpha_1\alpha_2}\rvert
  }{d_{\alpha_1}d_{\alpha_2}-\delta_{\alpha_1\alpha_2}}\right]
  $};

\end{scope}

\end{tikzpicture}%
}

\vspace{-2mm}

\caption{\label{fig:rtn_schematic}
Replica tensor network for the charge- and dipole-conserving random circuit.
The contraction gives the R\'enyi-2 purity-like quantities shown for $L=12$ and $L_A=3$.
Each integer time step applies one staggered layer of Haar-averaged four-site transfer
operators, shifted by one site relative to the previous step under OBCs. The lower
panel shows how Haar averaging of $(U^{*}\!\otimes U)^{\otimes2}$ gives the four-site
transfer tensor $\mathcal{T}_{i,\dots,i+3}$, subject to the appropriate charge and
dipole constraints. Here $\mathcal{O}$ is the untwisted swap for purity,
while $\mathcal{O}=Q$ and $P$ are the charge- and dipole-twisted swap
contractions for charge and dipole-resolved purity, respectively.}
\label{fig:schematic_RTN}
\end{figure}
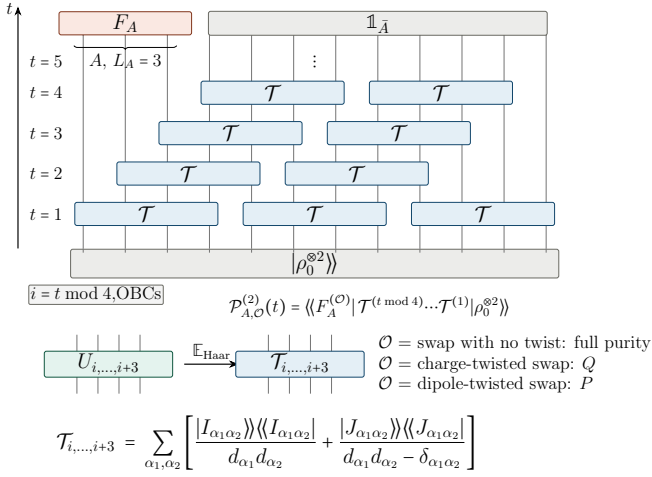

The R\'enyi-$2$ asymmetry involves purities of both the reduced density matrix and its symmetrized version, which are quadratic in $\rho$ and therefore require two replicas. We work in the Liouville (doubled) representation, in which an operator becomes a vector $|\rho\rangle\!\rangle$, unitary conjugation acts linearly, and purity is an overlap between the evolved replicated state and a swap boundary on the subsystem $A$. Each site thereby carries a \emph{doubled} space $\mathcal{H}_2\otimes\mathcal{H}_2$ (one factor for the ket and one for the bra index of $\rho$) with conjugation realized as $U\rho U^\dagger\mapsto(U\otimes U^{*})\,|\rho\rangle\!\rangle$. The two replicas required by the purity then promote this to the four-fold (``doublet'') space $(\mathcal{H}_2\otimes\mathcal{H}_2)^{\otimes2}$, i.e.\ two ket and two bra copies per site, on which the replicated gate $(U^{*}\!\otimes U)^{\otimes2}$ acts. Averaging this object over the Haar measure (performed independently in each charge- and dipole-symmetry block $\alpha$ of the four-site gate, of dimension $d_\alpha$) collapses it onto the local transfer tensor of Fig.~\ref{fig:schematic_RTN},
\begin{equation}\label{eq:transfer_tensor}
\mathcal{T}_{i:i+3}
=
\sum_{\alpha_1,\alpha_2}\bigg[
\frac{|I_{\alpha_1\alpha_2}\rangle\!\rangle\langle\!\langle I_{\alpha_1\alpha_2}|}{d_{\alpha_1}d_{\alpha_2}}
+
\frac{|J_{\alpha_1\alpha_2}\rangle\!\rangle\langle\!\langle J_{\alpha_1\alpha_2}|}{d_{\alpha_1}d_{\alpha_2}-\delta_{\alpha_1\alpha_2}}
\bigg].
\end{equation}
Here $\alpha_1,\alpha_2$ label the symmetry blocks of the two replicas, and $|I\rangle\!\rangle$, $|J\rangle\!\rangle$ are the only two structures that survive the second-moment (Weingarten) average: $|I_{\alpha_1\alpha_2}\rangle\!\rangle$ is the \emph{identity} pairing of the two replicas and $|J_{\alpha_1\alpha_2}\rangle\!\rangle$ the (orthogonalized) \emph{swap} pairing, weighted by the corresponding factors $1/(d_{\alpha_1}d_{\alpha_2})$ and $1/(d_{\alpha_1}d_{\alpha_2}-\delta_{\alpha_1\alpha_2})$. This is exactly the structure displayed in the lower panel of Fig.~\ref{fig:schematic_RTN}; the explicit construction and the reduced six-state per-site basis are given in Appendix~\ref{app:rtn_four_site}. Because the gates are independently drawn at each space-time position, the disorder average factorizes locally, and stacking these transfer tensors produces a two-dimensional tensor network (space along one direction, circuit time along the other). The bottom boundary is fixed by the replicated initial state $|\rho_0^{\otimes2}\rangle\!\rangle$, while the top boundary encodes the observable, as shown in Fig.~\ref{fig:schematic_RTN}.

The purity is then obtained by contracting this averaged
tensor with the standard swap boundary condition on $A$. By contrast, the charge- or dipole-symmetrized purities are obtained by replacing this boundary
condition with a symmetry-twisted swap. Importantly, the bulk tensor network
is the same in all cases. The distinction between the full purity, the
charge-symmetrized purity, and the dipole-symmetrized purity is entirely encoded in
the top boundary vector.

This boundary interpretation is especially useful. Symmetrization with
respect to $Q_A$ can be implemented as an average over charge rotations,
which in the replica tensor network dresses the swap boundary on $A$ by a
uniform phase twist. The dipole symmetrization has the same form, but the
twist is weighted by the site position because
$P_A=\sum_{i\in A} i\,q_i$. The corresponding boundary condition is therefore
site dependent: different sites in $A$ carry different phases. This
position-dependent twist is the replica tensor-network signature of the dipole symmetry. In this way, the entanglement asymmetries can be computed from $\mathcal P_A^{(2)}(t)$ and $\mathcal P_{A,\mathcal{O}}^{(2)}(t)$.
This formulation makes the computation efficient, because the same averaged replica transfer matrix can be reused for the different symmetry sectors.

\section{Dynamics in random circuits}
We open with the charge- and dipole-conserving random circuit, the setting in which the higher-order QME is sharpest and, through the replica tensor network, accessible at the largest scales. We first establish the strong Hilbert-space fragmentation of the model, and then show how it shapes the Mpemba-like crossings of the charge and dipole asymmetries, separating symmetry restoration into a frozen-memory and an active-relaxation channel.
\subsection{Hilbert-space fragmentation}
\begin{figure*}[!t]
\centering
\includegraphics[width=0.999\textwidth]{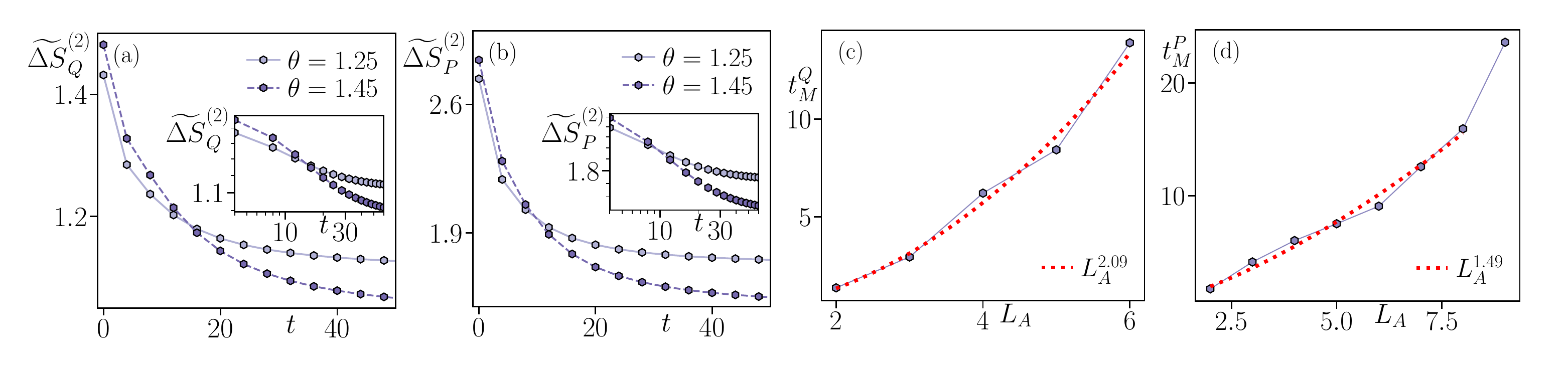}
\caption{
Charge and dipole asymmetry dynamics in a charge- and dipole-conserving Haar-random circuit with $L=128$.
(a-b) Charge and dipole second-R\'enyi asymmetries for tilted ferromagnetic initial states with $\theta=1.25$ and $\theta=1.45$, computed for a boundary subsystem of size $L_A=6$ using a replica tensor-network method with bond dimension $\chi=512$.
The curves exhibit a quantum-Mpemba-effect-like crossing, followed by slow late-time relaxation to a nonzero saturation value, as shown in the log-log insets.
(c-d) Crossing times $t_M^Q$ and $t_M^P$ vs $L_A$.
For small $L_A$, the data showcases $t_M^Q\propto L_A^{2.09}$ and $t_M^P\propto L_A^{1.49}$; for larger $L_A$, the crossings disappear within the accessible time window.
}
\label{fig:asymmetry_random_circuit}
\end{figure*}
We first present the numerical results for the charge- and dipole-conserving
spin-$\tfrac12$ circuit obtained with the replica tensor-network approach
(see Appendix~\ref{app:rtn_four_site} for the construction and Appendix~\ref{app:ED_RTN_benchmark} for the self-averaging of the asymmetry). The local dynamics is generated by the single allowed four-site move of Eq.~\eqref{eq:pairhop}, which preserves both the total charge $Q$ and the dipole moment $P$. This
simple constraint produces a highly fragmented Hilbert space. Since every gate is built from the single four-site exchange of Eq.~\eqref{eq:pairhop}, the computational basis states organize into disconnected orbits (the \emph{Krylov fragments}) that no sequence of gates can bridge: two configurations belong to the same fragment only if one is reachable from the other by repeated four-site exchanges. Most configurations admit no move at all and form \emph{frozen} fragments, single basis states left invariant by every gate that therefore never evolve, whereas the remaining \emph{active} fragments are extended orbits within which configurations freely interconvert and scramble. This split of the dynamics into frozen and active sectors is the structural fact behind everything that follows. As shown in
Appendix~\ref{app:fragment_counting}, the total number of Krylov fragments
grows exponentially, $N_{\rm frag}(L)\sim 1.839^L$, with an exponentially
large number of frozen single-configuration fragments,
$N_{\rm fr}(L)\sim 1.755^L$. The largest connected fragment,
\begin{equation}
D_{\max}(L)
=
\binom{\lfloor L/2\rfloor}{\lfloor L/4\rfloor}
\sim \frac{2^{L/2}}{\sqrt{\pi L/4}},
\end{equation}
is enormous in absolute terms, growing from $252$ states at $L=20$ to $\mathcal{O}(10^{18})$ at $L=128$, yet it occupies an exponentially vanishing fraction of its own charge-and-dipole sector,
\begin{equation}\label{eq:strong_frag_main}
\frac{D_{\max}(L)}{D_{0,0}(L)} \sim \frac{L^{3/2}}{2^{L/2}} \xrightarrow{\,L\to\infty\,} 0,
\qquad D_{0,0}(L)\sim \frac{2^L}{L^2},
\end{equation}
which is the defining criterion of \emph{strong} Hilbert-space fragmentation. The dynamics
therefore reflects a competition between frozen fragments, which preserve
exact memory of the initial state, and active fragments, which can relax
internally. Capturing this competition requires system sizes beyond exact
diagonalization, and is made possible by the replica tensor-network
framework.

\subsection{Higher-order quantum Mpemba effect}
The dynamics in Fig.~\ref{fig:asymmetry_random_circuit} reveals a
quantum-Mpemba-effect-like crossing in the R\'enyi-$2$ entanglement
asymmetries associated with both charge and dipole symmetrization, for a
subsystem attached to the boundary. For tilted ferromagnetic initial states
with $\theta=1.25$ and $\theta=1.45$, the state with initially larger
symmetry breaking relaxes faster and crosses below the initially less
asymmetric state at intermediate times. At later times, both asymmetries
relax very slowly, producing an apparent plateau over the numerically
accessible time window. The crossing-time analysis in
Fig.~\ref{fig:asymmetry_random_circuit}(c-d) further characterizes this
behavior: for small subsystems, the crossing times scale approximately as
$t_M^Q\sim L_A^{2.09}$ and $t_M^P\sim L_A^{1.49}$, while for larger
subsystems the crossing times diverge, indicating the disappearance of
observable crossings within the accessible time window.

\subsection{Krylov fragments: frozen memory and active relaxation}\label{sec:krylov}
Both features (the transient crossing and the late-time plateau) follow directly from the Krylov structure introduced above. A tilted product state is a superposition of computational configurations, so it spreads at once over many fragments, placing a fixed fraction of its weight in the frozen sector and the rest in the active sector. The two sectors then play opposite roles. On a frozen fragment the configuration is dynamically locked, and the reduced state keeps \emph{exact} memory of the initial symmetry breaking: its charge and dipole asymmetries never relax and set a finite floor that obstructs complete symmetry restoration, the late-time plateau seen in Fig.~\ref{fig:asymmetry_random_circuit}. On active fragments, by contrast, the four-site exchanges scramble the configurations and relax the asymmetry, and the more strongly tilted state relaxes faster, so its asymmetry crosses below that of the less tilted one, the Mpemba-like crossing.

\begin{figure*}[!t]
\includegraphics[width=0.99\textwidth]{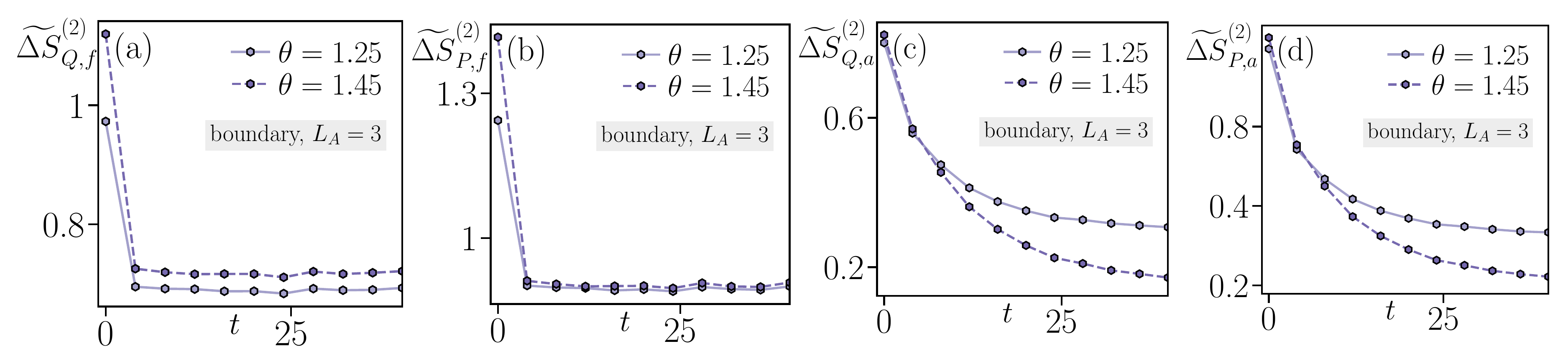}
\caption{\label{fig:frag_resolved}
Frozen and active contributions to the R\'enyi-$2$ entanglement asymmetries
for a boundary subsystem in the charge- and dipole-conserving random circuit.
The data are obtained for system size $L=16$ and subsystem size $L_A=3$,
starting from tilted ferromagnetic initial states with $\theta=1.25$ and
$\theta=1.45$. Panels (a,b) show the frozen-sector charge and dipole
asymmetries, $\Delta S^{(2)}_{Q,{\rm fr}}$ and
$\Delta S^{(2)}_{P,{\rm fr}}$, while panels (c,d) show the corresponding
active-sector asymmetries, $\Delta S^{(2)}_{Q,{\rm act}}$ and
$\Delta S^{(2)}_{P,{\rm act}}$. The frozen-sector asymmetries saturate at
nonzero plateaux and preserve the ordering between the two initial angles.
By contrast, the active-sector asymmetries decay in time and exhibit
Mpemba-like crossings, indicating that the transient acceleration of symmetry
restoration is associated with relaxation inside active Krylov fragments.}
\end{figure*}

We make this picture quantitative within each Krylov sector. Let $P_{\rm fr}=\sum_{\alpha\in{\rm frozen}}P_\alpha$ and $P_{\rm act}=\sum_{\alpha\in{\rm active}}P_\alpha$ project onto the frozen and active fragments, with $P_{\rm fr}+P_{\rm act}=\mathbb{1}$. Because the dynamics preserves each fragment, the weights $w_{\rm fr/act}=\langle\psi(t)|P_{\rm fr/act}|\psi(t)\rangle$ are time independent, and the diagonal blocks of $\rho(t)=|\psi(t)\rangle\langle\psi(t)|$ define normalized sector states $\tilde\rho_{\rm fr}=P_{\rm fr}\rho P_{\rm fr}/w_{\rm fr}$ and $\tilde\rho_{\rm act}=P_{\rm act}\rho P_{\rm act}/w_{\rm act}$. Tracing out the complement of $A$ gives $\rho_{A,{\rm fr/act}}(t)=\mathrm{Tr}_{\bar A}\,\tilde\rho_{\rm fr/act}(t)$, on which we evaluate the charge and dipole asymmetries separately. Because the asymmetry is nonlinear in $\rho_A$, this resolves the mechanism rather than providing an exact additive split, with the off-diagonal blocks $P_{\rm fr}\rho P_{\rm act}$ carrying the inter-sector coherence. Figure~\ref{fig:frag_resolved} confirms the two roles cleanly: the frozen-sector asymmetries saturate at nonzero plateaux, while the active-sector asymmetries decay and display the Mpemba-like crossings~\footnote{For a boundary subsystem the frozen-sector curves preserve the initial ordering of the two angles, whereas for a bulk subsystem they cross, reflecting the geometry dependence of the frozen-sector coherences probed by the partial trace.}.

The mechanism is more general than the model at hand: it rests only on the coexistence of frozen and active Krylov sectors, and not on the specific charge-and-dipole constraints. Hilbert-space fragmentation therefore does not preclude the higher-order QME but reshapes it, splitting symmetry restoration into a frozen channel that never completes and an active channel that drives the anomalous crossings.

In Appendix~\ref{app:bulk_asymmetry}, we also investigate bulk subsystems
using the same RTN approach. These data show similar QME crossings, but with
a smaller late-time saturation value than in the boundary case, as expected
from statistical edge localization
\cite{sala_ergo_2020,MAZUR1969533,SUZUKI1971277}.

We demonstrate the presence of the QME for another initial state, a tilted variant of the antiferromagnetic state, in Appendix~\ref{app:tilted_AFM}. Using exact vector simulation, we find qualitatively similar features to those observed for the tilted ferromagnetic state. This is in contrast with
standard $U(1)$-symmetric circuits, where the tilted antiferromagnetic state
does not display the same Mpemba-like behavior as the tilted ferromagnetic
state. Here, due to fragmentation, the QME appears in both cases.

\section{Beyond circuits}
Having established the higher-order QME and its fragmentation mechanism in the random circuit, we now test its robustness beyond the stochastic setting: first under coherent, energy-conserving Hamiltonian dynamics that share the same charge and dipole conservation, and then in an analytically solvable dissipative toy model.

\subsection{Hamiltonian dynamics}

\begin{figure*}
\includegraphics[width=0.99\textwidth]{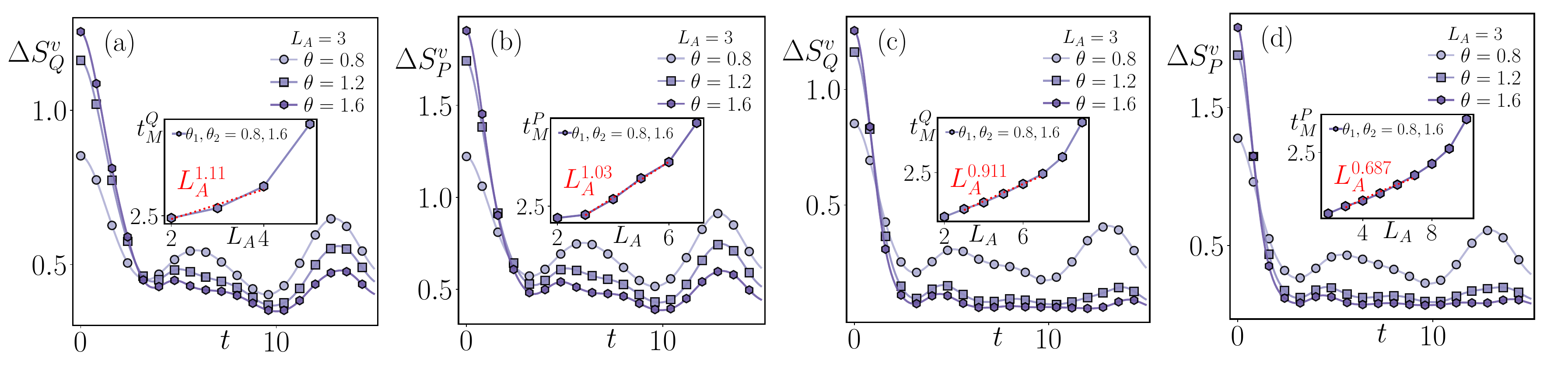}
\caption{Charge and dipole asymmetry dynamics for one-dimensional pair-hopping Hamiltonian of size $L=20$: (a-b): Von-Neumann asymmetries associated to charge and dipole for tilted ferromagnetic initial states with $\theta=0.8,1.2$ and $1.6$, computed for a boundary subsystem of size $L_A=3$. In both cases, we note characteristic QME-like crossings in both asymmetries with a finite late-time saturation value. (c-d): The same analysis for bulk subsystem of size $L_A=3$, which again showcases QME-crossings, with late-time dynamics showcasing the signature of symmetry restoration. In all four panels, we also showcase the crossing time vs $L_A$ behaviors in insets.  }
\label{fig:asym_ham}
\end{figure*}

We now focus the second setting: coherent
evolution under the pair-hopping
Hamiltonian~\eqref{eq:ham_dipole} with $L=20$
sites. Unlike the random circuit, where each gate is drawn independently and the entanglement asymmetry is computed using replica tensor networks, the Hamiltonian dynamics is
deterministic and we compute the von~Neumann
entanglement asymmetry
$\Delta S^{v}_{\mathcal{O}}$ using the Chebyshev
polynomial expansion~\cite{TalEzer1984AccurateEfficientScheme,Sierant2022ChallengesMBL}.
Figures~\ref{fig:asym_ham}(a--b) show the charge and dipole asymmetries for a subsystem of size $L_A=3$ attached to the boundary, starting from the tilted ferromagnetic states with $\theta=0.8$, $1.2$, and $1.6$. Both curves exhibit quantum-Mpemba-effect-like crossings: the state with
larger initial asymmetry relaxes faster than the other states, and the
corresponding curves cross at an intermediate time, before finite-size effects
become important. At later times, the asymmetries remain nonzero within the
accessible time window. While this residual value is reminiscent of the
plateau observed in the random circuit, in the Hamiltonian setting, it should
be interpreted with some care, since it may also be affected by finite-size
effects and by the limited time window accessible in exact simulations. The insets display the crossing time $t_M$ as a function of subsystem size.
Figures~\ref{fig:asym_ham}(c--d) repeat the same analysis for a subsystem of size $L_A=3$ detached from the boundary. The Mpemba-like crossings persist,
showing that the effect is not a boundary artifact. Compared with the boundary case, however, the late-time asymmetry is smaller, and the curves show
a less pronounced tendency to saturate. This suggests that the fragmentation-induced obstruction to symmetry restoration is weaker for a
bulk subsystem than for a subsystem attached to the boundary. Such a behavior
is consistent with the expectation that boundary subsystems are more strongly
affected by frozen configurations and statistical edge
localization~\cite{sala_ergo_2020,MAZUR1969533,SUZUKI1971277}, whereas bulk subsystems are less constrained by these boundary-localized frozen
structures.

In a nutshell, the Hamiltonian results prove that the higher-order Mpemba phenomenology identified in the stochastic circuit is not only due to the
random averaging: it arises from the underlying
charge and dipole conservation and persists under coherent, deterministic dynamics.

\subsection{Symmetric pair-flipping toy model}

We next consider the symmetric pair-flipping model, which provides an analytically tractable setting to isolate the interplay between dipole-conserving pair flips and local dephasing. The key simplification is that the dynamics factorizes into independent reflection-symmetric pairs. Indeed, for the Hamiltonian in Eq.~\eqref{eq:ham_sym} and local dephasing operators $L_x=S^z_x$, the Liouvillian can be
written as a sum of mutually commuting terms, $\mathcal L =\sum_{j=1}^{L/2}\mathcal L_{(j,L+1-j)}+\mathcal L_{\rm c}$, where the last term is present only for odd $L$. Thus each symmetric pair evolves independently, and the time-evolved density matrix factorizes into a product of two-site density matrices. This factorization allows us to compute the charged moments, and hence the Rényi-2 entanglement asymmetries, in closed form.
The analytical expressions are most conveniently written in terms of a small number of elementary building blocks, derived in the Appendix \ref{app:pair_factorized_dynamics}. Physically, these blocks correspond to the possible ways in which the subsystem $A$ intersects the symmetric pairs: $A$ can contain only one site of a symmetric pair, both sites, or, for odd $L$, the central site. The charged moments
are products of these elementary factors. If the charged moments become independent of $\alpha$, they reduce to the ordinary, uncharged moments, and the symmetry is restored. Conversely, if a residual $\alpha$-dependence persists at late times, the asymmetry remains finite. This makes the role of dephasing transparent. 

\begin{figure*}
\includegraphics[width=0.99\textwidth]{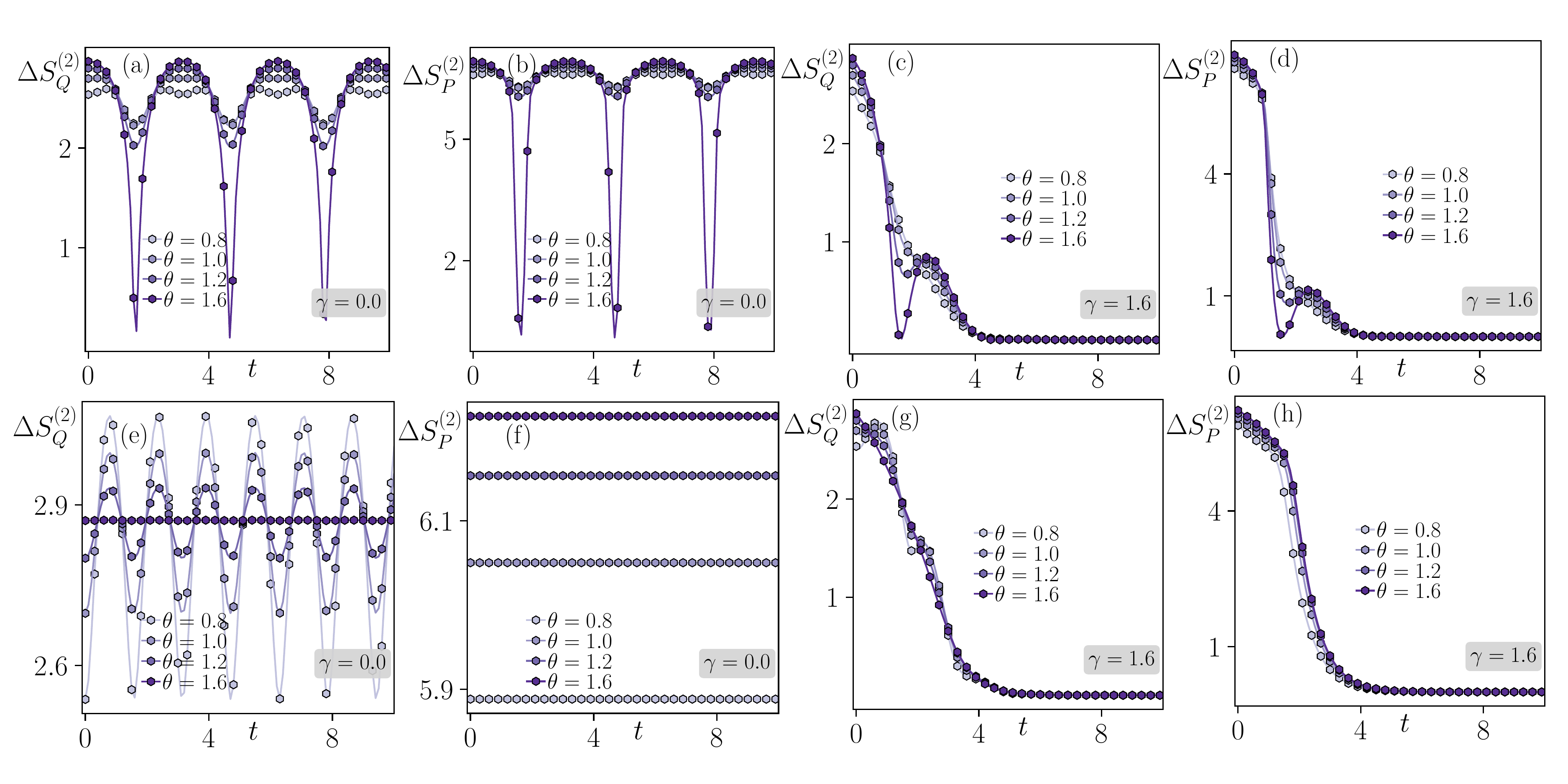}
\caption{
Charge- and dipole R\'enyi-2 asymmetries in the symmetric pair-flipping model for subsystems of size $L_A=100$. (a-d) Boundary geometry: $A$ consists of $L_A=100$ sites all on the same side of the chain center, $A=\{1,2,\dots,100\}$, so every site in $A$ has its mirror partner outside $A$. (e-h) Bulk geometry: $A$ is centered about the chain center and contains $L_A=99$ contiguous sites symmetric under reflection, so every mirror pair lies either entirely inside or entirely outside $A$ (the central site, if present, is included in $A$). (a,b,e,f) $\gamma=0$ (coherent dynamics): both asymmetries display persistent oscillatory or constant late-time behavior, with no signature of symmetry restoration. (c,d,g,h) $\gamma=1.6$ (dissipative, underdamped): both asymmetries decay, indicating restoration of the corresponding symmetries. In the boundary geometry (c,d) the asymmetries exhibit \emph{multiple Mpemba-like crossings} between curves with different $\theta$, as predicted by the closed-form $t_M$ of Eq.~\eqref{eq:tM_LA} and its periodic images; in the bulk geometry the charge sector (g) similarly shows crossings, while the dipole sector (h) shows \emph{no crossings}, the initial $\theta$-ordering being preserved throughout the relaxation.
}
\label{fig:pairflipping}

\end{figure*}

We probe the dynamics for two different geometries. We consider boundary and bulk subsystems of size $L_A=100$ and $99$, respectively, as shown in Fig. \ref{fig:pairflipping}. For the boundary geometry, $A=\{1,2,\dots,L_A\}$ contains one site from each of $L_A$ distinct mirror pairs and therefore probes $L_A$ \emph{independent} pair distances. By contrast, for the bulk geometry $A$ is centered about $j_0$ and contains $L_A$ contiguous sites symmetric under reflection. Each mirror pair therefore lies either entirely inside or entirely outside $A$.

For $\gamma=0$, the dynamics is purely coherent.
Consequently, neither the charge
nor the dipole symmetry is locally restored. Instead, both display persistent
oscillatory or nonzero late-time behavior, as shown in
Fig.~\ref{fig:pairflipping}(a-b,e-f). Thus dipole conservation alone
does not imply symmetry restoration.
The situation changes once local dephasing is added. For
\(\gamma=1.6\), the dynamics is underdamped but dissipative. The symmetric pairs can still oscillate, but the oscillations now acquire an
exponential envelope set by the dephasing rate. Hence, at late times, both charge and dipole asymmetries decay, as shown
in Fig.~\ref{fig:pairflipping}(c-d,g-h). In this sense, the dephasing
bath turns the coherent pair-flip dynamics into a genuine relaxation
problem while preserving the dipole-conserving structure.

The factorized dynamics of the symmetric pair-flipping model also explains
why the boundary geometry displays Mpemba crossings, and why the first
crossing time is essentially independent of the subsystem size. We focus on
the one-sided subsystem
\begin{equation}
A=\{1,2,\ldots,L_A\},
\end{equation}
which contains one site from each of $L_A$ distinct reflection-symmetric
pairs. This is the geometry used in Fig.~\ref{fig:pairflipping}(c-d), with
$L_A=100$. The solvability has a transparent origin. The Lindbladian of Eq.~\eqref{eq:lindblad} is a sum of terms supported on disjoint mirror pairs $(j_0{-}x,\,j_0{+}x)$ and on the central site, which mutually commute, so the density matrix factorizes at all times into independent two-site blocks. Within each pair the Hamiltonian connects only $\ket{+1/2,+1/2}\leftrightarrow\ket{-1/2,-1/2}$ (a single active two-level block) while $\ket{+1/2,-1/2}$ and $\ket{-1/2,+1/2}$ are frozen, and the local dephasing damps the surviving coherences at rate $\gamma$. Tracing each pair onto a single site then yields a charged moment of the simple form $g_O(\alpha)=\mathcal A(\theta,t)+\mathcal B(\theta,t)\cos\alpha$, whose Fourier resolution over the charge (or dipole) sectors gives the R\'enyi-2 asymmetry in closed form.
The exact solution derived in Appendix~\ref{app:pair_factorized_dynamics} shows that, in this geometry, the R\'enyi-$2$ asymmetry
reduces to $L_A$ identical late-time contributions, up to
subleading corrections. In the dissipative regime
$\gamma>0$, the oscillatory part of each contribution is exponentially
damped. This defines a small late-time parameter
\begin{equation}
\eta(\theta,t)
=
\frac{\mathcal B(\theta,t)}{\mathcal A(\theta,t)},
\end{equation}
where $\mathcal A$ is the non-oscillatory part of the single-pair
contribution and $\mathcal B$ is the damped oscillatory part as derived in Eqs. \eqref{eq:g_single_site_charge_explicit}, \eqref{eq:kappa_definition}, \eqref{eq:z_definition} as 
\begin{eqnarray}
\mathcal{A}&=&(1+z(t)^2)/2,~~~\mathcal{B}=2|\kappa(t)|^2,\nonumber\\
z(t)&=&e^{-\gamma t/2}\cos\theta\left[
\cos(2\Delta t)+\frac{\gamma/4}{\Delta}\sin(2\Delta t)
\right],\nonumber\\
\kappa(t)&=&\frac{\sin\theta}{2}e^{-\gamma t/2}
\left(\cos t+i\cos\theta\sin t\right).
\end{eqnarray}
Furthermore,
$\eta(\theta,t)$ vanishes exponentially at late times.

Expanding the exact result for the entanglement asymmetry in small $\eta$, one obtains
\begin{equation}
\Delta S^{(2)}_{O}(\theta,t)
=
L_A\,\eta(\theta,t)
+
O(L_A^2\eta^2),
\qquad
O\in\{Q,P\}.
\label{eq:DS_pairflip_leading_main}
\end{equation}
The leading term is the same for charge and dipole symmetry. It is also
linear in $L_A$, simply because the subsystem samples $L_A$ independent
reflection-symmetric pairs. The difference between the charge and dipole
sectors enters only through subleading corrections.
Therefore, at leading order, the crossing between two initial states
$\theta_1$ and $\theta_2$ is determined by
\begin{equation}
L_A\,\eta(\theta_1,t_M)
=
L_A\,\eta(\theta_2,t_M),
\end{equation}
thus the factor $L_A$ cancels. This explains why the first crossing time is
independent of the subsystem size and, to leading order, independent of
whether one probes charge or dipole symmetry.

To obtain the explicit crossing time, we use the dissipative late-time
approximation $\mathcal A(\theta,t_M)\simeq 1/2$. The crossing condition then
reduces to an equality between the damped oscillatory amplitudes,
\begin{equation}
\sin^2\theta_1
\left(
1-\sin^2\theta_1\sin^2 t_M
\right)
=
\sin^2\theta_2
\left(
1-\sin^2\theta_2\sin^2 t_M
\right),
\end{equation}
which gives
\begin{equation}
t_M
=
\arcsin
\left[
\frac{1}{\sqrt{\sin^2\theta_1+\sin^2\theta_2}}
\right].
\label{eq:tM_pairflip_main}
\end{equation}
A physical solution requires $\sin^2\theta_1+\sin^2\theta_2\geq 1$, thus the two initial states must break the symmetry strongly enough for the
ordering of the relaxation curves to invert.

This result captures the first Mpemba crossing observed in the boundary
geometry. Subleading terms depend on $L_A$ and on the specific operator $O$, but
they are suppressed when $L_A\eta(\theta,t)\lesssim 1$. For the parameters
used in Fig.~\ref{fig:pairflipping}, this condition is satisfied around the
first crossing, explaining why the analytical prediction remains accurate
even for the macroscopic subsystem $L_A=100$.
The same analysis also clarifies why more than one crossing can appear. The
single-pair contribution contains damped oscillatory functions of time, so
the equality $\eta(\theta_1,t)=\eta(\theta_2,t)$ can be satisfied multiple
times before the exponential envelope suppresses the oscillations. The first
solution is given by Eq.~\eqref{eq:tM_pairflip_main}, while later crossings
come from further solutions of the same exact factorized dynamics.
This gives a simple physical interpretation of the geometry
dependence. Subsystems that sample several independent symmetric pairs,
and hence several distance scales, combine multiple relaxation
channels and can favor Mpemba crossings more than other cases. 

The symmetric pair-flip model can be viewed as an analytically solvable realization of the same mechanism found in the random circuit. In the circuit, the Hilbert space
splits into frozen Krylov fragments, which retain memory of the
initial state, and active fragments, which help relaxation. In the
pair-flip model, each symmetric pair decomposes into two frozen one-dimensional
states, $\ket{+1/2,-1/2}$ and $\ket{-1/2,+1/2}$,
and one active two-dimensional subspace,
$\{\ket{+1/2,+1/2},\ket{-1/2,-1/2}\}$. The frozen
states play the role of the frozen Krylov fragments, while the active
two-level blocks provide the analogue of the active fragments. There is, however, an important difference. In the random circuit,
random gates generate relaxation inside the active fragments. In the
pair-flip model, the active blocks only undergo coherent oscillations
unless a dephasing bath is added. Thus dephasing
turns the active blocks into relaxing sectors. With dephasing, the active contribution can decay and speed up the symmetry restoration, while frozen components retain memory of the initial state. In Appendix-\ref{app:vonNeumann_pair_flipp}, we present exact numerical
results for the von Neumann entanglement asymmetry associated with charge and
dipole symmetry in the dissipative setting. These results show the same
qualitative features as the Rényi-2 asymmetry, confirming that the latter
provides a reliable probe of the QME.

\section{Summary and outlook}
\label{sec:conclusions}

We have shown that the quantum Mpemba effect not only survives strong Hilbert-space fragmentation, which splits the Hilbert space into exponentially many dynamically disconnected Krylov sectors, but acquires a \emph{higher-order} structure. We studied three complementary settings with both charge and dipole conservation: a Haar-random brickwork circuit, a pair-hopping Hamiltonian, and a dissipative pair-flip toy model, each highlighting a different aspect of the same phenomenology. In all three, the entanglement asymmetries associated with charge and dipole conservation display Mpemba-like crossings and relax on parametrically distinct timescales, one for each conserved quantity.

For the charge- and dipole-conserving random circuit with $L=128$, we compute
the dynamics using a replica tensor-network formulation adapted to four-site
dipole-conserving gates. For small subsystems, the crossing times scale as $t_M^{Q}\sim L_A^{2.09}$ for charge and $t_M^{P}\sim L_A^{1.49}$ for dipole,
before increasing sharply at larger $L_A$. Within the accessible time window,
both asymmetries also saturate to a nonzero plateau. An analysis of the different Krylov sectors explains these two features.
Frozen Krylov sectors produce a non-decaying asymmetry floor, which prevents
full symmetry restoration. By contrast, active sectors allow relaxation and
are responsible for the Mpemba-like crossings. The same physical picture also appears in the pair-hopping Hamiltonian, despite
the dynamics now being deterministic and coherent rather than random. In this
case, we also observe a difference between boundary and bulk subsystems:
boundary subsystems show more pronounced plateaux and are more strongly
affected by frozen sectors, while bulk subsystems show weaker plateaux and
clearer signatures of symmetry restoration. Finally, we study a dissipative pair-flipping model whose Liouvillian splits
into independent reflection-symmetric pairs. This simple structure allows us
to obtain analytical results for the higher-order QME. For suitable
subsystem geometries, we find multiple Mpemba-like crossings. The earliest crossing time is $t_M=
\arcsin\!(
1/\sqrt{\sin^2\theta_1+\sin^2\theta_2})$, which is independent of both the subsystem size $L_A$ and the choice of
symmetry, $\mathcal O\in\{Q,P\}$. The model also shows that dephasing is necessary. When $\gamma=0$, the
exponential factor $e^{-\gamma t}$ is absent, so the curves may cross
repeatedly, but they do not show a
monotonic relaxation toward symmetry restoration.

 Several directions naturally extend the present work. 
 
 (i) \emph{Higher conserved moments.} The framework developed here applies, in principle, to any constrained model whose local moves are compatible with a hierarchy of conserved quantities; quadrupole-conserving and more generally $k$-moment-conserving circuits would offer a natural testbed for a tower of independent Mpemba effects~\cite{gotta2026enhancingentanglementasymmetryfragmented}, each associated with a different conservation law and potentially relaxing on its own timescale. 
 
 (ii) \emph{Higher dimensions.} Fracton-like dipole conservation in two and higher dimensions~\cite{Will24HSFtwodimension,gromov24fractonhydrodynamics}, where the kinematic constraints on local moves are even more stringent and the fragmentation structure correspondingly richer, would test the generality of the frozen-plus-active decomposition identified here. 
 
 (iii) \emph{Quantitative theory of the plateau.} Although separating frozen and active sectors, as done in Sec.~\ref{sec:krylov}, provides a qualitative attribution of the late-time plateau to the frozen sector, a quantitative prediction of its dependence on $\theta$, $L_A$, and subsystem geometry is currently missing. A comprehensive analytical theory to underpin the mechanism might therefore be an interesting future direction. 
 
 (iv) \emph{QME and fragmentation-induced freezing transition.} Another exciting follow-up would be to understand how constraint-induced strong-weak fragmentation transition~\cite{Morningstar_2020,abhisodh2024,East_Sreemayee,Aditya25higherEast,Feldmeier21criticallyslow} can alter the QME in these systems.

 (v)\emph{The interplay between QME, fragmentation, and open quantum systems} Another follow-up question would be to examine the interplay between QME in open quantum systems where dephasing can reduce the quantum-to-classical fragmentation transition~\cite{Li23openHSF}. 
 
 (vi)\emph{Experimental probes.} The higher-order QME identified here is sharp enough at the R\'enyi-2 level (which is experimentally accessible through randomized measurements and shadow ~\cite{Brydges2019Renyirandomized,Elben2020randomizedmixedstate,Elben2023randomizedmeasurement,joshi-24,Rath23entanglementbarrier}) to be observable in dipole-conserving various experimental platforms platforms~\cite{wang25HSFexp,Adler2024FragmentationFractons2D}, where both charge and dipole conservation can be engineered through tilted-lattice protocols.  
 
 (vii) \emph{Beyond entanglement asymmetry.} It would be interesting to ask whether other quantum resource measures~\cite{Chitambar2019,aditya2025growthspreadingquantumresources}, such as non-stabilizerness~\cite{leone2022stabilizer,leone2024stabilizer,Liu22magic,turkeshi2025magic}, coherence~\cite{luitz,saxena20coherence,Baumgratz14coherence,lami,Turkeshi24Delocalization,Magni2025AnticoncentrationClifford,SauliereUniversality2025,MagniTurkeshi2025QuantumComplexity,magni2025anticoncentrationstatedesigndoped} and non-Gaussianity~\cite{Takagi18NG,sierant2025fermionicmagicresourcesquantum,ares2026asymmetrylowerboundfermionic,ares2026nongaussianityrandomquantumstates,santra2025quantumresourcesnonabelianlattice,tobia,lumia,w97w-7zny,pedro} exhibit fragmentation-induced versions of the Mpemba phenomenology, an avenue that we leave for future work.

More broadly, the frozen-plus-active structure suggests using the higher-order QME as a probe of how constrained quantum devices store and erase information: frozen fragments protect a long-lived symmetry-breaking memory (the plateaux) while active fragments relax it (the crossings), a coexistence recently observed in Rydberg arrays~\cite{Datla2025pyv}. Preparing families of initial states with tunable symmetry breaking and reading out the R\'enyi-2 asymmetry would then discriminate memory protected by frozen sectors from memory that thermalizes through active ones, turning the entanglement asymmetry into a direct diagnostic of memory lifetimes and relaxation bottlenecks in Rydberg and trapped-ion simulators.

\begin{acknowledgments}
S.A. gratefully acknowledges Piotr Sierant for previous collaborations on related topics and for sharing valuable insights on numerous occasions into the development and optimization of several numerical simulation codes, which have also been greatly beneficial for this work. S.A. and S.M. also thank Pasquale Calabrese, Filiberto Ares, and Riccardo Senese for helpful discussions on the quantum Mpemba effect, and S.A. acknowledges Diptiman Sen, Deepak Dhar, and Maitri Ganguli for previous collaborations on Hilbert-space fragmentation. S.A. further acknowledges support from the Alexander von Humboldt Foundation through a Humboldt Postdoctoral Fellowship.
X.T. acknowledges support from DFG Emmy Noether Programme proposal ``\textit{Digital Quantum Matter Ouf-of-Equilibrium}'' No. 560726973, DFG under Germany's Excellence Strategy – Cluster of Excellence Matter and Light for Quantum Computing (ML4Q) EXC 2004/2 – 390534769, and DFG Collaborative Research Center (CRC) 183 Project No. 277101999 - project B01.

\textbf{Data and Code Availability.---} Data and code will be publicly shared upon publication.
\end{acknowledgments}

\appendix

\onecolumngrid

\section{Replica tensor network for four-site charge- and dipole-conserving gates}
\label{app:rtn_four_site}

This appendix details the replica tensor-network construction that yields the disorder-averaged R\'enyi-2 entanglement asymmetries of the charge- and dipole-conserving random circuit. The local Hilbert space is $\mathcal{H}_2 \simeq \mathbb{C}^2$ (spin-$\tfrac{1}{2}$, $q=2$), with computational basis $\{|x\rangle \equiv |\sigma\rangle\}$, $\sigma = 0, 1$. Throughout we work in the Liouville (doubled) representation introduced in the main text: an operator $A$ becomes a vector $|A\rangle\!\rangle$, with $\mathrm{tr}(B^\dagger A)=\langle\!\langle B|A\rangle\!\rangle$ and $U A U^\dagger\mapsto(U\otimes U^{*})\,|A\rangle\!\rangle$, so that the R\'enyi-2 purities live on the two-replica (``doublet'') space.

As discussed
in the main text, the key computational ingredient is
the Haar average of the fourth moment of each four-site
gate. Since the gates at different space-time positions
are drawn independently, the average factorizes into a
product of local transfer operators. The basic building
block is
\[
T_{i,i+1,i+2,i+3}
= \mathbb{E}_{\mathrm{Haar}}\!\big[
  (U_{i,i+1,i+2,i+3}^{*} \otimes U_{i,i+1,i+2,i+3})^{\otimes 2}
\big],
\]
which acts on the four-replica space of four contiguous
sites. This object replaces the random gate with an
effective deterministic transfer matrix that encodes
all circuit-averaged correlations.

The charge-
and dipole-conserving gate decomposes into independent
symmetry sectors labelled by $\alpha = (Q, P)$, where
$Q$ is the total charge and $P$ the total dipole moment
across the four sites
[cf.\ Eq.~\eqref{eq:gate} of the main text]. Within
each sector the gate acts as an independent Haar-random
unitary of dimension $d_\alpha \equiv d_{Q,P} = \dim\mathcal{H}_{Q,P}$.
The transfer operator inherits this block structure: the
Haar average is performed independently in each sector,
and the cross-sector terms factorize. Explicitly, the
transfer operator splits as

\begin{eqnarray}
T_{i:i+3} &=& \sum_{\alpha_1 \neq \alpha_2} \Big[
\mathbb{E}_{\mathrm{Haar}}\big(
  U_{i:i+3}^{(\alpha_1)*} \otimes U_{i:i+3}^{(\alpha_2)}
  \otimes U_{i:i+3}^{(\alpha_2)*} \otimes U_{i:i+3}^{(\alpha_1)}
\big)
+
\mathbb{E}_{\mathrm{Haar}}\big(
  U_{i:i+3}^{(\alpha_1)*} \otimes U_{i:i+3}^{(\alpha_1)}
  \otimes U_{i:i+3}^{(\alpha_2)*} \otimes U_{i:i+3}^{(\alpha_2)}
\big)
\Big] \nonumber \\
&&+ \sum_{\alpha}
\mathbb{E}_{\mathrm{Haar}}\big(
  U_{i:i+3}^{(\alpha)*} \otimes U_{i:i+3}^{(\alpha)}
  \otimes U_{i:i+3}^{(\alpha)*} \otimes U_{i:i+3}^{(\alpha)}
\big),
\label{eq:T_decomp}
\end{eqnarray}
where $U^{(\alpha)}$ denotes the restriction of the gate
to the symmetry sector $\alpha = (Q, P)$, the first sum
runs over pairs of \emph{distinct} sectors
$\alpha_1 \neq \alpha_2$, and the second sum runs over
the diagonal $\alpha_1 = \alpha_2 = \alpha$. 
To express the
result of the Haar average, we introduce two families
of invariant states in the four-replica space, labelled
by pairs of symmetry sectors $(\alpha_1, \alpha_2)$.
The ``identity'' and "swap" pairing on four-site blocks can thus be written as
\begin{eqnarray}
|I^{+}_{\alpha_1,\alpha_2}\rangle\!\rangle
&=& \sum_{\{x^{(r)}\}} \bigotimes_{r=1}^{4}
|x^{(r)}_1\, x^{(r)}_1\, x^{(r)}_2\, x^{(r)}_2\rangle\!\rangle\;
\delta\!\Big(\sum_{r=1}^{4} \alpha_r^{(1)}, \alpha_1\Big) \delta\!\Big(\sum_{r=1}^{4} \alpha_r^{(2)}, \alpha_2\Big),
\nonumber\\
|I^{-}_{\alpha_1,\alpha_2}\rangle\!\rangle
&=& \sum_{\{x^{(r)}\}} \bigotimes_{r=1}^{4}
|x^{(r)}_1\, x^{(r)}_2\, x^{(r)}_2\, x^{(r)}_1\rangle\!\rangle\;
\delta\!\Big(\sum_{r=1}^{4} \alpha_r^{(1)}, \alpha_1\Big) \delta\!\Big(\sum_{r=1}^{4} \alpha_r^{(2)}, \alpha_2\Big),
\end{eqnarray}
where $\alpha = (Q, P)$, and the Kronecker delta enforces
that the total charge and total dipole moment of the
four-site configuration match the sector labels; the
addition of $\alpha$'s is understood component-wise
(charge added to charge, dipole added to dipole).
Applying
the Haar fourth-moment identity within each symmetry
sector, the three types of contributions in
Eq.~\eqref{eq:T_decomp} evaluate to:
\begin{eqnarray}
\mathbb{E}_{\mathrm{Haar}}(
  U_{\alpha_1}^* \otimes U_{\alpha_2}
  \otimes U_{\alpha_2}^* \otimes U_{\alpha_1})
&=& \frac{1}{d_{\alpha_1}\, d_{\alpha_2}}\;
|I^{-}_{\alpha_1,\alpha_2}\rangle\!\rangle
\langle\!\langle I^{-}_{\alpha_1,\alpha_2}|,
\nonumber\\[4pt]
\mathbb{E}_{\mathrm{Haar}}(
  U_{\alpha_1}^* \otimes U_{\alpha_1}
  \otimes U_{\alpha_2}^* \otimes U_{\alpha_2})
&=& \frac{1}{d_{\alpha_1}\, d_{\alpha_2}}\;
|I^{+}_{\alpha_1,\alpha_2}\rangle\!\rangle
\langle\!\langle I^{+}_{\alpha_1,\alpha_2}|,
\nonumber\\[4pt]
\mathbb{E}_{\mathrm{Haar}}(
  U_{\alpha}^* \otimes U_{\alpha}
  \otimes U_{\alpha}^* \otimes U_{\alpha})
&=& \frac{1}{d_\alpha^2 - 1}\bigg[
|I^{+}_{\alpha,\alpha}\rangle\!\rangle
\langle\!\langle I^{+}_{\alpha,\alpha}|
+ |I^{-}_{\alpha,\alpha}\rangle\!\rangle
\langle\!\langle I^{-}_{\alpha,\alpha}|
- \frac{1}{d_\alpha}\Big(
  |I^{+}_{\alpha,\alpha}\rangle\!\rangle
  \langle\!\langle I^{-}_{\alpha,\alpha}|
  + |I^{-}_{\alpha,\alpha}\rangle\!\rangle
  \langle\!\langle I^{+}_{\alpha,\alpha}|
\Big)\bigg].
\end{eqnarray}
The first two lines correspond to the cross-sector
($\alpha_1 \neq \alpha_2$) terms, where the two
replicas probe independent Haar-random unitaries and the
result is simply a product of first moments. The third
line is the same-sector ($\alpha_1 = \alpha_2 = \alpha$)
contribution, which involves the full Weingarten
function for $U(d_\alpha)$ at $n = 2$ and produces
the characteristic cross-terms between the identity
and swap pairings. The
Haar-averaged transfer operator can be brought into
a particularly transparent form by introducing the
orthogonalised states
\[
|J_{\alpha_1,\alpha_2}\rangle\!\rangle
\equiv
|I^{-}_{\alpha_1,\alpha_2}\rangle\!\rangle
- \delta_{\alpha_1,\alpha_2}\,
\frac{1}{d_{\alpha_1}}\,
|I^{+}_{\alpha_1,\alpha_2}\rangle\!\rangle,
\]
and renaming
$|I^{+}_{\alpha_1,\alpha_2}\rangle\!\rangle
\to |I_{\alpha_1,\alpha_2}\rangle\!\rangle$.
With this notation, the full Haar-averaged gate takes
the compact form
\begin{eqnarray}
\overline{U^* \otimes U \otimes U^* \otimes U}
&=&
\sum_{\alpha_1,\alpha_2}
\frac{1}{d_{\alpha_1}\, d_{\alpha_2}}\;
|I_{\alpha_1,\alpha_2}\rangle\!\rangle
\langle\!\langle I_{\alpha_1,\alpha_2}|
+\sum_{\alpha_1,\alpha_2}
\frac{1}{d_{\alpha_1}\, d_{\alpha_2} - \delta_{\alpha_1,\alpha_2}}\;
|J_{\alpha_1,\alpha_2}\rangle\!\rangle
\langle\!\langle J_{\alpha_1,\alpha_2}|,
\nonumber\\
\label{eq:compact_T}
\end{eqnarray}
which makes manifest the two-channel structure of the
replica average: the $|I\rangle\!\rangle$ channel
corresponds to the identity pairing across replicas,
while the $|J\rangle\!\rangle$ channel encodes the
swap pairing with the appropriate orthogonalization.
The sums now run over all pairs $(\alpha_1, \alpha_2)$
including the diagonal, and the Kronecker delta in the
denominator of the second term automatically
distinguishes the same-sector and cross-sector
contributions.

For spin-$\tfrac{1}{2}$, the na\"ive per-site replica
space is eight-dimensional, spanned by the states
$|\pm,\, r,\, b\rangle\!\rangle_i$ with $r, b = 0, 1$ (corresponding to $\sigma_{i}=\pm 1/2$).
However, two exact degeneracies reduce this to six
linearly independent states:
\begin{equation}
|{+},\, 1,\, 1\rangle\!\rangle_i
=
|{-},\, 1,\, 1\rangle\!\rangle_i,
\quad
|{+},\, 0,\, 0\rangle\!\rangle_i
=
|{-},\, 0,\, 0\rangle\!\rangle_i.
\label{eq:degeneracies}
\end{equation}
The effective per-site replica space is therefore
$C_i^{\mathrm{eff}} = \mathrm{span}\{|\pm, r, b\rangle\!\rangle_i
\mid r, b = 0, 1\}$
subject to these identifications, yielding six
independent states per site. 
Therefore, averaging over the Haar ensemble of four-site charge and dipole conserving case simplifies
to the 1296-dimensional space generated by the on-site replica basis $C_{i}^{{\rm eff}}$.
We now
connect the transfer operator to the physical
observables. Starting from a pure initial state
$|\Psi_0\rangle$, the system evolves to
$\rho(t) = U_t\, |\Psi_0\rangle\langle\Psi_0|\, U_t^\dagger$
after $t$ time steps. The R\'enyi-2 purity of the
reduced density matrix on a subsystem $A$ is
\begin{equation}
P_A(t) = \mathrm{Tr}\!\big(
  F_A\, [\rho(t) \otimes \rho(t)]
\big),
\end{equation}
where $F_A = \bigotimes_{i \in A} F_i$ is the swap
operator acting on subsystem $A$, with local action
$F(|x\rangle \otimes |y\rangle) = |y\rangle \otimes |x\rangle$.
Passing to the Liouville representation
$\rho \mapsto |\rho\rangle\!\rangle$, the unitary
conjugation becomes the linear map
$(U_t^* \otimes U_t)|\rho_0\rangle\!\rangle$, and the
purity takes the form
\begin{equation}
P_A(t) = \langle\!\langle F_A|\,
(U_t^* \otimes U_t \otimes U_t^* \otimes U_t)\,
|\rho_0^{\otimes 2}\rangle\!\rangle.
\end{equation}
Since the local gates at each space-time position are
independent and identically distributed (i.i.d.),
the circuit average can be performed layer by layer.
Denoting the full averaged transfer matrix for one
time step as $\mathcal{T}$ (obtained by composing the
four staggered layers of local transfer operators),
the disorder-averaged purity becomes
\begin{equation}
\mathcal{P}_A^{(2)}(t)
= \mathbb{E}[P_A(t)]
= \langle\!\langle F_A|\,
\mathcal{T}^t\,
|\rho_0^{\otimes 2}\rangle\!\rangle.
\label{eq:avg_purity}
\end{equation}
This expression has a natural interpretation as the
contraction of a two-dimensional tensor network: the
spatial direction corresponds to the chain sites, the
temporal direction to the circuit layers, the bottom
boundary is set by the initial state
$|\rho_0^{\otimes 2}\rangle\!\rangle$, and the top
boundary by the swap operator
$\langle\!\langle F_A|$.

To compute the entanglement asymmetries,
we need the charge-dephased and dipole-dephased
purities, which require inserting appropriate
projectors into the trace. We introduce two dephasing
channels acting on subsystem $A$. The \emph{charge-dephasing channel} projects onto
eigenspaces of the subsystem charge operator
$Q_A = \sum_{i \in A} q_i$:
\begin{equation}
\mathcal{C}_A(X) \equiv \sum_{q_A} \Pi_{q_A}\, X\, \Pi_{q_A},
\end{equation}
where $\Pi_{q_A}$ is the projector onto the eigenspace
with charge $q_A$. For a subsystem of $N_A$
spin-$\tfrac{1}{2}$ sites, the charge takes
$N_A + 1$ values ($q_A = 0, 1, \ldots, N_A$). The
projectors can be resolved via a discrete Fourier
transform by introducing the angles
$\phi_k = 2\pi k / (N_A + 1)$ with $k = 0, \ldots, N_A$,
which yields the Fourier representation of the projectors,
\begin{equation}
\Pi_{q_A} = \frac{1}{N_A + 1}
\sum_{k=0}^{N_A} e^{-i\phi_k q_A}\, e^{i\phi_k Q_A},
\end{equation}
and leads to the twirling form of the channel:
\begin{equation}
\mathcal{C}_A(X)
= \frac{1}{N_A + 1}
\sum_{k=0}^{N_A} e^{-i\phi_k Q_A}\, X\, e^{i\phi_k Q_A}.
\label{eq:charge_twirl}
\end{equation}
This representation is computationally advantageous
because the twirling operators $e^{i\phi_k Q_A}$
factorize over sites, allowing the dephasing to be
absorbed into the local boundary conditions of the
tensor network.

Analogously, the \emph{dipole-dephasing channel}
projects onto eigenspaces of the subsystem dipole
operator $P_A = \sum_{i \in A} r_i\, q_i$:
\begin{equation}
\mathcal{D}_A(X) \equiv \sum_{p_A} \Pi_{p_A}\, X\, \Pi_{p_A}.
\end{equation}
Introducing Fourier angles
$\theta_\ell = 2\pi\ell / N_P$ (with $N_P$ the number
of distinct dipole eigenvalues), the analogous
twirling form reads
\begin{equation}
\mathcal{D}_A(X)
= \frac{1}{N_P}
\sum_{\ell=0}^{N_P - 1}
e^{-i\theta_\ell P_A}\, X\, e^{i\theta_\ell P_A}.
\end{equation}

The charge-dephased and dipole-dephased reduced
density matrices are then
\begin{equation}
\rho_{A,Q_A}(t) \equiv \mathcal{C}_A(\rho_A(t)),
\qquad
\rho_{A,P_A}(t) \equiv \mathcal{D}_A(\rho_A(t)),
\end{equation}
with corresponding R\'enyi-2 purities
\begin{equation}
P^{(2)}_{A,Q_A}(t)
= \mathrm{Tr}\!\big[\rho_{A,Q_A}(t)^2\big],
\qquad
P^{(2)}_{A,P_A}(t)
= \mathrm{Tr}\!\big[\rho_{A,P_A}(t)^2\big].
\end{equation}

In the two-replica Liouville representation, these
purities can be written as
\begin{align}
P^{(2)}_{A,Q_A}(t) &= \langle\!\langle F_A|\,
(\mathcal{C}_A \otimes \mathcal{C}_A)\,
|\rho(t) \otimes \rho(t)\rangle\!\rangle,
\\
P^{(2)}_{A,P_A}(t) &= \langle\!\langle F_A|\,
(\mathcal{D}_A \otimes \mathcal{D}_A)\,
|\rho(t) \otimes \rho(t)\rangle\!\rangle.
\end{align}
A key
simplification arises from the fact that both
$\mathcal{C}_A$ and $\mathcal{D}_A$ are completely
positive trace-preserving maps constructed from
orthogonal projectors. They are therefore self-adjoint
with respect to the Hilbert--Schmidt inner product,
$\mathrm{Tr}[X^\dagger\,\mathcal{C}_A(Y)]
=\mathrm{Tr}[\mathcal{C}_A(X)^\dagger\,Y]$, and
\emph{idempotent}, $\mathcal{C}_A^2=\mathcal{C}_A$ (since
$\Pi_{q_A}\Pi_{q_A'}=\delta_{q_A q_A'}\Pi_{q_A}$), and
likewise for $\mathcal{D}_A$. These two properties let us
twist a \emph{single} replica rather than both. Using
self-adjointness and idempotence together with the
Hermiticity of $\rho_A$,
\begin{equation}
\mathrm{Tr}\!\big[\mathcal{C}_A(\rho_A)^2\big]
=\mathrm{Tr}\!\big[\mathcal{C}_A^2(\rho_A)\,\rho_A\big]
=\mathrm{Tr}\!\big[\mathcal{C}_A(\rho_A)\,\rho_A\big],
\label{eq:idempotent_collapse}
\end{equation}
so the symmetrized purity needs the channel on one copy
only, $(\mathcal{C}_A\otimes\mathbb{1})$, and the
dephasing can be moved entirely onto the boundary vector,
defining the \emph{twisted swap states}
\begin{equation}
|F'_A\rangle\!\rangle
\equiv (\mathcal{C}_A \otimes \mathbb{1})\,
|F_A\rangle\!\rangle,
\qquad
|F''_A\rangle\!\rangle
\equiv (\mathcal{D}_A \otimes \mathbb{1})\,
|F_A\rangle\!\rangle,
\end{equation}
so that the dephased purities become
\begin{equation}
P^{(2)}_{A,Q_A}(t)
= \langle\!\langle F'_A|\,
|\rho(t) \otimes \rho(t)\rangle\!\rangle,~~~
P^{(2)}_{A,P_A}(t)
= \langle\!\langle F''_A|\,
|\rho(t) \otimes \rho(t)\rangle\!\rangle.
\end{equation}
Thus, the bulk of the tensor
network (determined by the Haar-averaged gates) is
\emph{identical} for all three purities
($P_A$, $P^{(2)}_{A,Q_A}$, and $P^{(2)}_{A,P_A}$).
The only difference lies in the choice of boundary
vector at the top of the network: the bare swap
$|F_A\rangle\!\rangle$ for the full purity, and the
charge- or dipole-twisted swap for the dephased
purities. This separation of bulk and boundary is
what makes the computation efficient. Using the twirling representations of the dephasing
channels, the twisted boundary states take the
explicit Fourier form
\begin{align}
|F_A^{(Q)}\rangle\!\rangle
&= \frac{1}{N_Q}
\sum_{k=0}^{N_Q-1} |F_A(\phi_k)\rangle\!\rangle,
\\
|F_A^{(P)}\rangle\!\rangle
&= \frac{1}{N_P}
\sum_{\ell=0}^{N_P-1} |F_A(\theta_\ell)\rangle\!\rangle,
\end{align}
where, the Fourier angles are
$\phi_k = 2\pi k / N_Q$ with $k = 0, \ldots, N_Q - 1$
and $N_Q = N_A + 1$, since a subsystem of $N_A$
spin-$\tfrac{1}{2}$ sites admits $N_A + 1$ distinct
charge eigenvalues $q_A = 0, 1, \ldots, N_A$.
Similarly, $\theta_\ell = 2\pi\ell / N_P$ with
$\ell = 0, \ldots, N_P - 1$, where $N_P$ counts the
number of distinct subsystem dipole eigenvalues.
For the coordinate convention $r_i = 1, 2, \ldots, N_A$,
the subsystem dipole $P_A = \sum_{i \in A} r_i\, q_i$
ranges from $p_A^{\min} = 0$ to
$p_A^{\max} = \sum_{i=1}^{N_A} i = N_A(N_A+1)/2$,
giving $N_P = N_A(N_A+1)/2 + 1$ distinct sectors. As both
$Q_A$ and $P_A$ are sums of single-site operators,
the twisted boundary states factorize over sites
within $A$:
\begin{align}
|F_i(\phi_k)\rangle\!\rangle
&= \big(e^{-i\phi_k q_i} \otimes e^{i\phi_k q_i}\big)\otimes\mathbb{1}\,
|F_i\rangle\!\rangle,
\\
|F_i(\theta_\ell)\rangle\!\rangle
&= \big(e^{-i\theta_\ell r_i q_i} \otimes e^{i\theta_\ell r_i q_i}\big)\otimes\mathbb{1}\,
|F_i\rangle\!\rangle,
\end{align}

Since
the local twisted swap states $|F_i(\phi_k)\rangle\!\rangle$
and $|F_i(\theta_\ell)\rangle\!\rangle$ live inside the
six-dimensional effective space
$C_i^{\mathrm{eff}}$ identified in
Eq.~\eqref{eq:degeneracies}, each can be expanded in
the effective replica basis
$|\!-,\,r,\,b\rangle\!\rangle_i$ with $r, b = 0, 1$ as
\begin{equation}
|F_i(\phi_k)\rangle\!\rangle
= \sum_{r,b=0,1}
A_k(r,b)\;
|\!-,\, r,\, b\rangle\!\rangle_i\,,
\end{equation}
with expansion coefficients $A_k(r,b) = e^{i\phi_k(r - b)}$. For the dipole-twisted states, the same structure holds
with the replacement $\phi_k \to \theta_\ell r_i$,
giving
\begin{equation}
|F_i(\theta_\ell)\rangle\!\rangle
= \sum_{r,b=0,1}
A_\ell^{(i)}(r,b)\;
|\!-,\, r,\, b\rangle\!\rangle_i\,,
\end{equation}
where $A_\ell^{(i)}(r,b) = e^{i\theta_\ell r_i(r - b)}$.
In contrast to the charge case, the dipole coefficients
$A_\ell^{(i)}(r,b)$ depend on the site coordinate $r_i$
through the product $\theta_\ell r_i$ in the exponent.
Each site within $A$ therefore carries a distinct local
boundary condition. This site-dependence is the
fingerprint of the higher-order (dipole) symmetry in
the replica framework, and constitutes the key
structural difference from the standard
charge-only construction.

For
the tilted ferromagnet [Eq.~\eqref{eq:tilted_ferro}], each
site carries
$|\psi_i(\theta)\rangle
= \cos\!\tfrac{\theta}{2}\,|0\rangle
+ \sin\!\tfrac{\theta}{2}\,|1\rangle$,
and the two-replica initial vector factorises as
$|\rho_0^{\otimes 2}\rangle\!\rangle
= \bigotimes_i \sum_{\sigma=\pm}\sum_{r,b=0,1}
B_\theta^\sigma(r,b)\,|\sigma,r,b\rangle\!\rangle_i$
with $B_\theta^+ = B_\theta^- \equiv B_\theta$, where
\begin{equation}
B_\theta(r,b)
= \big(\cos\tfrac{\theta}{2}\big)^{2(2-r-b)}\,
\big(\sin\tfrac{\theta}{2}\big)^{2(r+b)}\,.
\end{equation}

With help of these, the annealed-averaged
charge and dipole-R\'enyi-2 purities
are computed by contracting the \emph{same} bulk
replica tensor network, built from the Haar-averaged
four-site transfer operators of
Eq.~\eqref{eq:compact_T}, against different top
boundary conditions, computable via tensor network methods for system sizes larger than
exact computation techniques permit.

\section{Self-averaging of the entanglement asymmetry}
\label{app:ED_RTN_benchmark}

The replica tensor network, and the main text, work with the \emph{annealed} asymmetry, a ratio of circuit-averaged purities, whereas the Mpemba criterion is conceptually phrased in terms of the \emph{typical} (quenched) asymmetry of a single realization. The two coincide only if the entanglement asymmetry is \emph{self-averaging}. The sole purpose of this appendix is to check, using exact vector simulation (ED), how closely the annealed average tracks the quenched value at the parameters used in the main text.
The replica tensor-network method introduced in Appendix~\ref{app:rtn_four_site} computes the \emph{annealed} R\'enyi-2 entanglement asymmetries of Eq.~\eqref{eq:annealed_QP}, defined as the logarithm of a ratio of circuit-averaged purities,
\begin{equation}
\widetilde{\Delta S}^{(2)}_{A,O}(t) \;=\; -\,\log\!\frac{\mathbb{E}[\,\mathrm{tr}\,\rho_{A,O}^2\,]}{\mathbb{E}[\,\mathrm{tr}\,\rho_{A}^2\,]} \,,
\qquad O\in\{Q,P\}\,.
\label{eq:annealed_def_app}
\end{equation}
The same quantity is in principle accessible to exact vector simulation (ED) by sampling the two purities separately and averaging each before taking the logarithm. By contrast, the another quantity one can naturally extract from the exact vector simulation is the \emph{quenched} R\'enyi-2 asymmetry,
\begin{equation}
\Delta S^{(2)}_{A,O}(t) \;=\; \mathbb{E}\!\left[\,-\log\!\frac{\mathrm{tr}\,\rho_{A,O}^2}{\mathrm{tr}\,\rho_{A}^2}\,\right] \,,
\label{eq:quenched_def_app}
\end{equation}
in which the asymmetry is computed for each individual realization and only then averaged over the Haar ensemble. By Jensen's inequality, the two quantities satisfy $\widetilde{\Delta S}^{(2)}_{A,O}(t) \le \Delta S^{(2)}_{A,O}(t)$, with equality if and only if the fluctuations of the purity ratio across realisations are negligible.

Concretely, we compare the annealed and quenched averages of Eqs.~\eqref{eq:annealed_def_app}--\eqref{eq:quenched_def_app}, both computed within ED from the same set of trajectories, and show that the Jensen-inequality gap between them is numerically negligible at the parameters of interest, so that the qualitative dynamical features (the Mpemba-like crossings, the $\theta$-ordering of the curves, and the late-time plateau) are properties of the typical realization, not artifacts of the annealed prescription used in the main text. We consider a charge- and dipole-conserving brickwall random circuit of size $L=20$, evolved from tilted ferromagnetic initial states with $\theta=1.25$ and $\theta=1.45$, and compute both the boundary and bulk R\'enyi-2 asymmetries on a subsystem of size $L_A=3$. The ED data are obtained by sampling $4000$ Haar-random gate configurations and evolving the state exactly under each realization; both the annealed and quenched asymmetries of Eqs.~\eqref{eq:annealed_def_app}--\eqref{eq:quenched_def_app} are then constructed from the same set of trajectories.

\begin{figure*}
\includegraphics[width=0.95\textwidth]{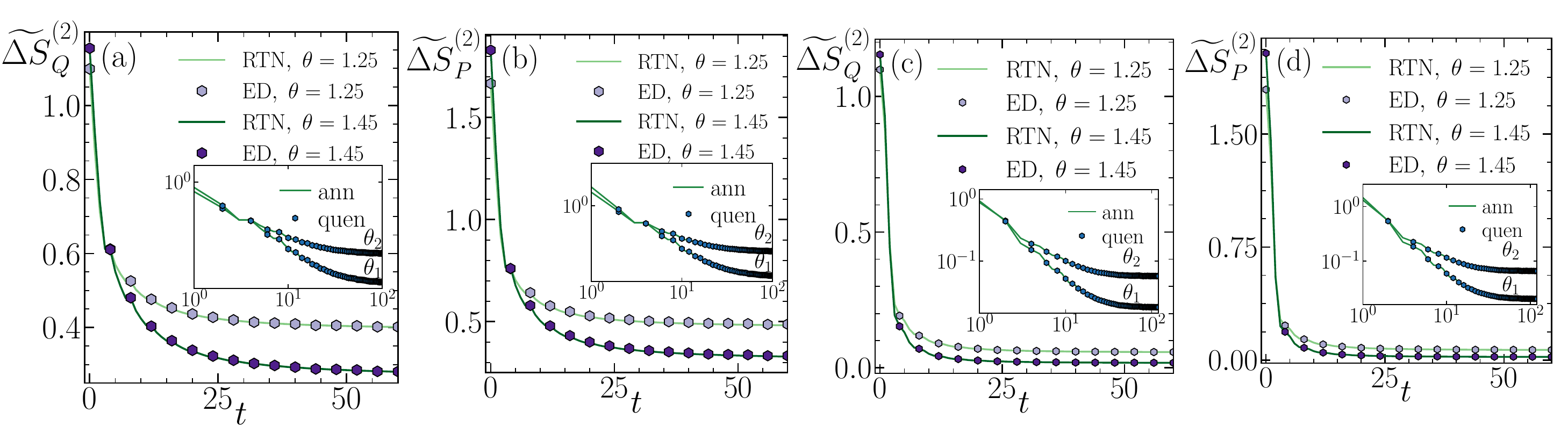}
\caption{Self-averaging of the entanglement asymmetry in the charge- and dipole-conserving Haar-random circuit of size $L=20$, with tilted ferromagnetic initial states at $\theta=1.25$ and $\theta=1.45$.
(a-b) R\'enyi-2 charge and dipole asymmetries for a \emph{boundary} subsystem of size $L_A=3$.
(c-d) Same quantities for a \emph{bulk} subsystem of size $L_A=3$.
\emph{Main panels:} the annealed R\'enyi-2 asymmetry (lines: replica tensor network at $\chi=512$; symbols: exact vector simulation over at least $4000$ realizations).
\emph{Log-log insets:} the same data on a logarithmic time axis, together with the quenched ED average of Eq.~\eqref{eq:quenched_def_app}, plotted separately for $\theta_1=1.25$ and $\theta_2=1.45$. For each tilting angle, the annealed (``ann'') and quenched (``quen'') curves overlap to within the resolution of the data}
\label{fig:ED_RTN_benchmark}
\end{figure*}

The results are collected in Fig.~\ref{fig:ED_RTN_benchmark}. The main panels (a-d) show the annealed asymmetry $\widetilde{\Delta S}^{(2)}_{A,O}(t)$ for charge and dipole, in the boundary and bulk geometries. The decisive comparison is in the log-log insets, where for each tilting angle the annealed (``ann'') average is overlaid on the quenched (``quen'') ED average of Eq.~\eqref{eq:quenched_def_app}: the two are numerically indistinguishable over the entire time window, so the per-realization fluctuations of the purity ratio are small enough that the Jensen gap $\widetilde{\Delta S}^{(2)}_{A,O}\le \Delta S^{(2)}_{A,O}$ is invisible. The entanglement asymmetry therefore self-averages at these parameters, and the annealed prescription used in the main text faithfully reproduces the quenched asymmetry on which the QME criterion is conceptually based.

\section{Fragmentation structure of charge and dipole-conserving spin-1/2 circuits}
\label{app:fragment_counting}

This appendix derives the fragment counts quoted in the main text. Writing $1\equiv +1/2$ and $0\equiv -1/2$, the Krylov fragments are the orbits of the four-site move $1001 \leftrightarrow 0110$ permitted by the gate of Eq.~\eqref{eq:gate}, and frozen states are those containing neither $1001$ nor $0110$; all counts below follow from exact enumeration, independently validated with transfer-matrix methods.

\subsection{Total number of fragments}

Let $N_{\rm frag}(L)$ denotes the total number of
Krylov fragments for a chain of length~$L$ in OBCs. To count $N_{\rm frag}$
analytically, we construct a transfer matrix~
\cite{Dhar_1993,menon_1997,HariMenon_1995,Deepak_HSF,East_Sreemayee,Aditya25higherEast} on the
eight three-site configurations
$\{000,001,010,011,100,101,110,111\}$.
The matrix element $T_{\rm frag}(c_i,c_j)$ connects
the row state $(s_j,s_{j+1},s_{j+2})$ to the column
state $(s_{j+1},s_{j+2},s_{j+3})$ whenever the
overlap is consistent and the stitched four-site
string does \emph{not} equal $1100$. The resulting $8\times 8$
matrix thus reads as
\begin{equation}
T_{\rm frag}=
\begin{pmatrix}
1&1&0&0&0&0&0&0\\
0&0&1&1&0&0&0&0\\
0&0&0&0&1&1&0&0\\
0&0&0&0&0&0&1&1\\
1&1&0&0&0&0&0&0\\
0&0&1&1&0&0&0&0\\
0&0&0&0&0&1&0&0\\
0&0&0&0&0&0&1&1
\end{pmatrix}
\label{eq:T_frag_app}
\end{equation}
and the total number of fragments can be computed as\cite{Dhar_1993,menon_1997,HariMenon_1995,Deepak_HSF,East_Sreemayee,Aditya25higherEast}
\begin{equation}
N_{\rm frag}(L)=\sum_{i,j=1}^{8}
\left(T_{\rm frag}^{\,L-3}\right)_{ij},
\qquad L\ge 4.
\label{eq:Nfrag_tm_app}
\end{equation}
whose values are shown in Table-\ref{tab:frag_counts}, and exactly agrees the values obtained using exact basis enumeration method. The largest eigenvalue of $T_{\rm frag}$ is
$\lambda_{\rm frag}^{\max}=1.839$, thus leading to growth of
$N_{\rm frag}$ as $1.839^L$ asymptotically and $\lambda_{\rm frag}^{\max}$ is the largest root of $x^3=x^2+x+1$, the tribonacci constant. Taking the initial values $N_{\rm frag}(L)
= 1,2,4,8$ for $L=0,1,2,3$
one can also verify that $N_{\rm frag}$ satisfies the
recursion relation
\begin{equation}
N_{\rm frag}(L) = 2\,N_{\rm frag}(L-1)
- N_{\rm frag}(L-4),
\label{eq:Nfrag_recursion}
\end{equation}
equivalently $N_{\rm frag}(L)=\sum_{k\le L}T_k$ with $T_k$ the tribonacci numbers.

\begin{table}[t]
\begin{ruledtabular}
\begin{tabular}{ccccccccc}
$L$ & 4 & 5 & 6 & 7 & 8 & 9 & 10 & 11 \\
\hline
$N_{\rm frag}$ & 15 & 28 & 52 & 96 & 177 & 326 & 600 & 1104 \\
\end{tabular}
\end{ruledtabular}
\caption{\label{tab:frag_counts}
Total number of fragments $N_{\rm frag}(L)$ for open
boundary conditions, obtained from BFS enumeration
and verified against the transfer matrix in
Eq.~\eqref{eq:T_frag_app}.}
\end{table}

\subsection{Frozen fragments}

In classically fragmented systems, a basis configuration is frozen if and only if it
contains neither $1001$ nor $0110$ on any contiguous
four-site block. The corresponding transfer matrix
$T_{\rm fr}$ is constructed in the same way as
$T_{\rm frag}$, but now \emph{both} patterns are
forbidden:
\begin{equation}
T_{\rm fr}=
\begin{pmatrix}
1&1&0&0&0&0&0&0\\
0&0&1&1&0&0&0&0\\
0&0&0&0&1&1&0&0\\
0&0&0&0&0&0&0&1\\
1&0&0&0&0&0&0&0\\
0&0&1&1&0&0&0&0\\
0&0&0&0&1&1&0&0\\
0&0&0&0&0&0&1&1
\end{pmatrix}
\label{eq:T_frozen_app}
\end{equation}
yielding \cite{Dhar_1993,menon_1997,HariMenon_1995,Deepak_HSF,East_Sreemayee,Aditya25higherEast}
\begin{equation}
N_{\rm fr}(L)=\sum_{i,j=1}^{8}
\left(T_{\rm fr}^{\,L-3}\right)_{ij},
\qquad L\ge 4,
\label{eq:Nfrozen_tm_app}
\end{equation}
whose values are shown in Table-\ref{tab:froz_counts}.
The largest eigenvalue of $T_{fr}$ is
$\lambda_{\rm fr}^{\max}=1.755$, thus leading to the growth of frozen
states $1.755^{L}$, slower than the growth of $N_{frag}$. Again, setting $N_{\rm fr}$=1,2,4,8 for $L$=0,1,2,3, this sequence satisfies
the recursion relation
\begin{equation}
N_{\rm fr}(L) = 2\,N_{\rm fr}(L{-}1)
- N_{\rm fr}(L{-}2) + N_{\rm fr}(L{-}3).
\label{eq:Nfr_recursion}
\end{equation}

\begin{table}[t]
\begin{ruledtabular}
\begin{tabular}{ccccccccc}
$L$ & 4 & 5 & 6 & 7 & 8 & 9 & 10 & 11 \\
\hline
$N_{\rm fr}$ & 14 & 24 & 42 & 74 & 130 & 228 & 400 & 702 \\
\end{tabular}
\end{ruledtabular}
\caption{
Number of frozen states $N_{\rm fr}(L)$ for open
boundary conditions, obtained from BFS enumeration
and verified against the transfer matrix in
Eq.~\eqref{eq:T_frozen_app}.}
\label{tab:froz_counts}
\end{table}
\subsection{Dimension of largest fragment and strong fragmentation}

The dimension of the largest fragment can be obtained
analytically by mapping the local move to a
permutation problem. Grouping the chain into two-site
blocks and defining $10\to A$ and $01\to B$, the
local move $1001\leftrightarrow 0110$ becomes
$AB\leftrightarrow BA$. Within a fully connected
fragment the dynamics therefore simply permutes a
word of $n_A$ letters $A$ and $n_B$ letters $B$,
giving the fragment dimension
$D(n_A,n_B)=\binom{n_A+n_B}{n_A}$.
The largest fragment corresponds to the most balanced
composition with $m=\lfloor L/2\rfloor$ blocks,
\begin{equation}
D_{\max}(L)=
\binom{\lfloor L/2 \rfloor}
{\left\lfloor \lfloor L/2 \rfloor/2 \right\rfloor},
\label{eq:Dmax_app}
\end{equation}
whose values are listed in
Table~\ref{tab:largest_fragment}. Asymptotically,
$D_{\max}(L)\sim 2^{L/2}/\sqrt{\pi L/4}$.

To diagnose the nature of fragmentation, the relevant
comparison is growth of the largest fragment compared to the dimension of conventional symmetry resolved sector, i.e., the dimension $D_{0,0}(L)$ with $(Q{=}0,P{=}0)$ (in which largest fragment resides for $L=4m$). The counting of $D_{0,0}$ is equivalent to
counting the partitions of $\{1,2,\ldots,L\}$ into two
equal-size subsets with equal sums, which is nonzero
only for $L\equiv 0\pmod{4}$ (since $Q=0$ forces
$L$ even, and $P=0$ further requires the target
sum $L(L+1)/4$ to be an integer).

Although no closed-form expression for $D_{0,0}$ is
known, it can be extracted as a coefficient of a
constrained generating function. Introducing a
fugacity $y$ that tracks the number of spin-up
($s_i=+\tfrac{1}{2}$) sites and a variable $x$
conjugate to the dipole moment, one has
\begin{equation}
D_{0,0}(L) = \big[x^{L(L+1)/4}\,y^{L/2}\big]
\prod_{j=1}^{L}(1+x^j\,y)\,,
\label{eq:D00_genfun}
\end{equation}
where $[x^a y^b]f(x,y)$ denotes the coefficient
of $x^a y^b$ in the Taylor expansion of $f$.
The constraint $y^{L/2}$ enforces $Q=0$ (equal
number of $+\tfrac{1}{2}$ and $-\tfrac{1}{2}$ spins),
while $x^{L(L+1)/4}$ enforces $P=0$ (the sum of
site indices carrying $s_i=+\tfrac{1}{2}$ equals
$L(L+1)/4$). Equivalently, $D_{0,0}$
admits the Fourier integral representation
\begin{equation}
D_{0,0}(L) = \frac{1}{(2\pi)^2}
\int_0^{2\pi}\!\!\int_0^{2\pi}
\prod_{j=1}^{L}
\big(1+e^{i(\alpha + j\beta)}\big)\,
d\alpha\,d\beta\,,
\label{eq:D00_fourier}
\end{equation}
where $\alpha$ and $\beta$ are the Fourier
conjugates to the charge and dipole constraints,
respectively. The exact values obtained from
Eq.~\eqref{eq:D00_genfun} are listed in
Table~\ref{tab:D00_values}. A saddle-point
analysis of Eq.~\eqref{eq:D00_fourier} in the
large-$L$ limit yields the asymptotic scaling
\begin{equation}
D_{0,0}(L) \sim \frac{c\, 2^L}{L^2}\,,
\label{eq:D00_asymptotic}
\end{equation}
with $c=O(1)$, where the $L^{-2}$ suppression
relative to $\binom{L}{L/2}\sim 2^L/\sqrt{L}$
reflects the additional dipole constraint beyond
charge conservation alone. The ratio of the largest
fragment to this sector dimension therefore satisfies
\begin{equation}
\frac{D_{\max}(L)}{D_{0,0}(L)}
\sim \frac{L^{3/2}}{2^{L/2}}
\;\xrightarrow{L\to\infty}\; 0\,,
\label{eq:strong_frag_criterion}
\end{equation}
which vanishes exponentially. This establishes that
the charge- and dipole-conserving spin-$\tfrac12$
circuit exhibits \emph{strong} Hilbert-space
fragmentation~:
even within the $(Q{=}0,P{=}0)$ symmetry sector, the
largest dynamically connected fragment occupies an
exponentially vanishing fraction of the available
states.

\begin{table}[t]

\begin{ruledtabular}
\begin{tabular}{ccccccccc}
$L$ & 4 & 5 & 6 & 7 & 8 & 9 & 10 & 11 \\
\hline
$D_{\max}$ & 2 & 2 & 3 & 3 & 6 & 6 & 10 & 10 \\
\end{tabular}
\end{ruledtabular}
\caption{\label{tab:largest_fragment}
Dimension $D_{\max}(L)$ of the largest fragment from
Eq.~\eqref{eq:Dmax_app}, verified against exact basis enumeration method.}
\end{table}

\begin{table}[t]

\begin{ruledtabular}
\begin{tabular}{cccc}
$L$ & $D_{0,0}(L)$ & $D_{\max}(L)$ &
$D_{\max}/D_{0,0}$ \\
\hline
4   & 2     & 2    & 1.000 \\
8   & 8     & 6    & 0.750 \\
12  & 58    & 20   & 0.345 \\
16  & 526   & 70   & 0.133 \\
20  & 5448  & 252  & 0.046 \\
\end{tabular}
\end{ruledtabular}
\caption{\label{tab:D00_values}
Dimension $D_{0,0}(L)$ of the $(Q{=}0,P{=}0)$
symmetry sector and the ratio $D_{\max}/D_{0,0}$,
for $L\equiv 0\pmod{4}$.}
\end{table}
\section{Asymmetry dynamics for bulk subsystems}
\label{app:bulk_asymmetry}

Fragmented systems exhibit pronounced boundary effects (notably statistical edge localization~\cite{sala_ergo_2020}) that can obscure the bulk physics. Here we verify that the features reported in the main text for boundary subsystems persist when $A$ sits deep in the bulk of the chain.

\begin{figure*}
\includegraphics[width=1.0\textwidth]{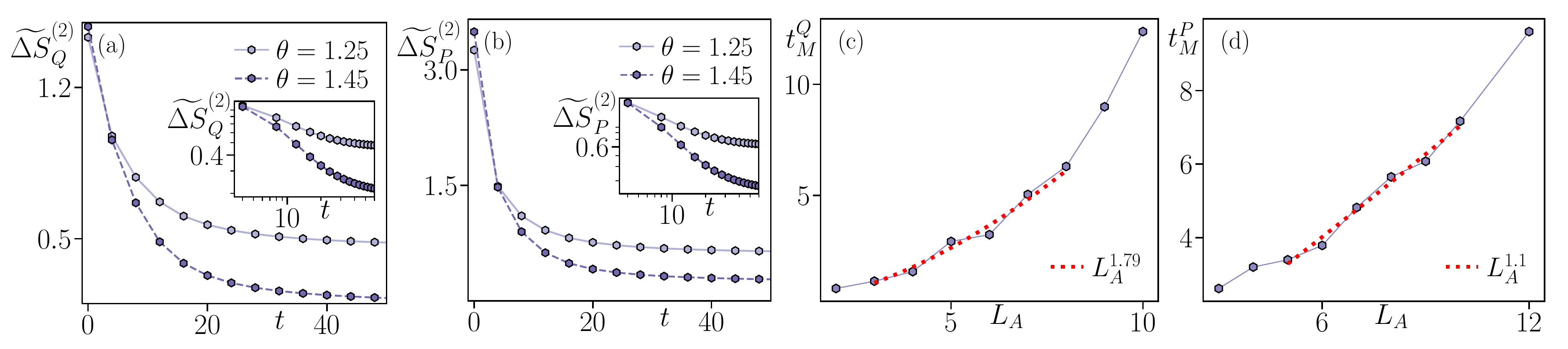}
\caption{Charge and dipole asymmetry dynamics in a charge- and dipole-conserving Haar-random circuit with $L=128$.
(a-b) Charge- and dipole second-R\'enyi asymmetries for tilted ferromagnetic initial states with $\theta=1.25$ and $\theta=1.45$, computed for a bulk subsystem of size $L_A=6$ using a replica tensor-network method with $\chi=512$.
The curves exhibit a quantum-Mpemba-effect-like crossing, as for boundary subsystems, followed by slow late-time relaxation to a nonzero saturation value, as shown in the log-log insets.
(c-d) Crossing times $t_M^Q$ and $t_M^P$ vs $L_A$, which show features similar to the boundary subsystem.}
\label{fig:QME_circuit_bulk}
\end{figure*}

Figures~\ref{fig:QME_circuit_bulk}(a--b) show the
Rényi-2 charge and dipole asymmetries,
$\Delta S^{(2)}_{A,Q}$ and $\Delta S^{(2)}_{A,P}$,
for the charge- and dipole-conserving random circuit
with $L=128$ sites and a bulk subsystem of size
$L_A=6$, starting from tilted ferromagnetic initial
states with $\theta=1.25$ and $\theta=1.45$. Both
asymmetries display the same phenomenology observed
for boundary subsystems: (i) Mpemba-like crossings,
where the initially more asymmetric state
($\theta=1.45$) restores symmetry faster than the
less asymmetric one ($\theta=1.25$); (ii) slow
relaxation consistent with the anomalous dynamics
induced by fragmentation; and (iii) saturation at a
nonzero value within the accessible time window, with
The plateau value is smaller than in the boundary
case, reflecting the weaker influence of the frozen
sector in the bulk.
The Mpemba crossing times $\tau_M^Q$ and $\tau_M^P$
scale with subsystem size as
$\tau_M^Q \propto L_A^{1.79}$ and
$\tau_M^P \propto L_A^{1.1}$, qualitatively
consistent with the boundary scaling reported in the
main text. However, the crossing time beyond a specific $L_A$ increases abruptly similar to the boundary case, indicating the absence of Mpemba-like crossing for larger $L_A$'s.

\section{QME for tilted antiferromagnetic states in charge- and dipole-conserving spin-$\tfrac{1}{2}$ circuits}
\label{app:tilted_AFM}

\begin{figure*}
\includegraphics[width=1.0\textwidth]{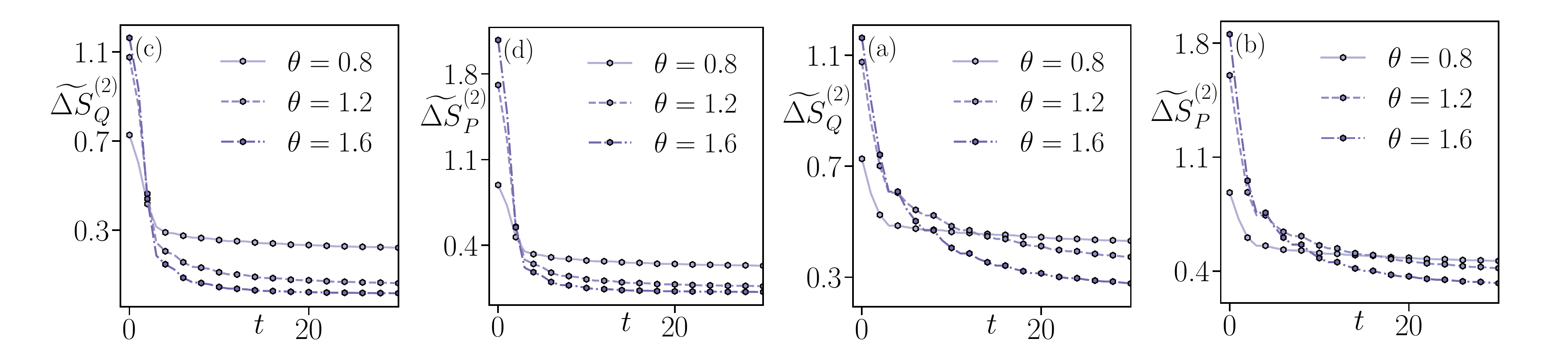}
\caption{Charge and dipole asymmetry dynamics for charge and dipole-conserving circuit of size $L=20$ using exact vector simulation: (a-b):  Charge and dipole annealed averaged Renyi-2 asymmetries for tilted Anti-ferromagnetic initial states with $\theta=0.8,1.2$ and $1.6$, computed for a bulk subsystem of size $L_A=3$. In both cases, we note characteristic QME-like crossings in both asymmetries with the amount of symmetry restoration depending on the value of $\theta$-value. (c-d): The same analysis for boundary subsystem of size $L_A=3$, which again showcases QME-crossings, with late time saturation value being much larger than the bulk case. In all cases, the data is appropriately averaged over at least 4000 circuit realizations.}
\label{fig:tilted_AFM}
\end{figure*}

The main text uses the tilted ferromagnetic state as the prototypical symmetry-breaking initial condition. To test how strongly the higher-order QME depends on this choice, we repeat the analysis for the tilted antiferromagnetic (AFM) state
\begin{equation}
|\Psi_0^{\rm AFM}(\theta)\rangle = \bigotimes_{i=1}^{L} e^{-i(-1)^{i+1}Y_i\theta/2}\,|s_i\rangle,\qquad
|s_i\rangle = \begin{cases}\ket{+1/2}, & i\ \text{odd},\\[2pt]
\ket{-1/2}, & i\ \text{even},\end{cases}
\label{eq:tilted_AFM}
\end{equation}
which is the staggered tilt of a classical N\'eel configuration.  In Figure~\ref{fig:tilted_AFM}, we showcase the charge- and dipole-resolved annealed R\'enyi-2 entanglement asymmetries for $L=20$ and a subsystem of size $L_A=3$, evolved under the same charge- and dipole-conserving brickwall random circuit considered in the main text, with $\theta\in\{0.8,1.2,1.6\}$ and all the quantities are averaged over at least $4000$ random circuit realizations.

The AFM state exhibits the same higher-order QME phenomenology as the FM state: in both the bulk [Fig.~\ref{fig:tilted_AFM}(a,b)] and boundary [Fig.~\ref{fig:tilted_AFM}(c,d)] geometries, and in both the charge and dipole sectors, the more strongly tilted states relax faster and cross the less tilted ones at intermediate times. This is in marked contrast to $U(1)$-conserving random circuits, where the QME is absent for tilted antiferromagnetic initial conditions.

The bulk and boundary reduced dynamics display qualitatively distinct late-time behaviours that mirror those identified for the FM state in the main text and in Appendix~\ref{app:bulk_asymmetry}. In the bulk geometry, the curves at the larger tilting angle ($\theta=1.6$) decay toward a small residual value, indicating effective symmetry restoration within the accessible time window, while the curves at smaller tilting ($\theta\lesssim 1.2$) exhibit a clear plateau. The natural interpretation, parallel to the fragment-resolved analysis of Sec.~\ref{sec:krylov}, follows from the projection of the initial state onto the frozen Krylov sector: the unrotated N\'eel configuration $\ket{+1/2,-1/2,+1/2,-1/2,\cdots}$ is itself a frozen state, so for small $\theta$ the tilted-AFM state retains a large overlap with the frozen sector, which is then directly inherited by the late-time density matrix and produces a pronounced asymmetry plateau. As $\theta$ increases towards $\pi/2$ the tilted state spreads more uniformly over the computational basis, the frozen-sector weight decreases, and a larger fraction of the state becomes available to the thermalizing active fragments, thus leading to symmetry restoration. 

Furthermore, the boundary geometry shows a substantially more pronounced lack of symmetry restoration than the bulk geometry: the late-time saturation values in Fig.~\ref{fig:tilted_AFM}(c,d) are markedly larger than those in panels (a,b), and the curves for all $\theta$ values remain well separated from zero throughout the simulation window. This bulk-boundary asymmetry is consistent with the statistical edge-localization phenomenology of fragmented systems, already seen for the tilted ferromagnetic state in the main text. Together, these observations demonstrate that the higher-order QME survives a change of initial state from FM to AFM.

\section{Analytical results for third setup}

\label{app:pair_factorized_dynamics}

\subsection{Derivation of asymmetries}

We consider an odd chain of length $L=2m+1$, with sites paired by reflection about the central site. The coherent dynamics is generated by the mirror-pair Hamiltonian
\begin{equation}
H=\sum_{j=1}^{M}\left(S_j^+S_{L+1-j}^+ + S_j^-S_{L+1-j}^-\right),
\label{eq:mirror_hamiltonian}
\end{equation}
and the dissipative dynamics is due to local dephasing jumps
\begin{equation}
L_i=S_i^z,\qquad i=1,\ldots,L .
\label{eq:dephasing_jumps}
\end{equation}
The density matrix evolves according to
\begin{eqnarray}
\partial_t\rho&=& -i[H,\rho]
+\gamma\sum_{i=1}^{L}(
L_i\rho L_i^\dagger -\frac{1}{2}\{L_i^\dagger L_i,\rho\}
).
\label{eq:lindblad_equation}
\end{eqnarray}

The key structural property of this model is that the Liouvillian decomposes into mutually commuting contributions acting on disjoint spatial supports, namely
\begin{equation}
\mathcal L=\sum_{j=1}^{M}\mathcal L_{(j,L+1-j)}+\mathcal L_{\rm c},
\label{eq:liouvillian_decomposition}
\end{equation}
where $\mathcal L_{(j,L+1-j)}$ acts only on the mirror pair $(j,L+1-j)$, and $\mathcal L_{\rm c}$ acts only on the central site. Since the local terms have disjoint support, they commute. Consequently, for any initial product state over sites, the time-evolved density matrix remains factorized into independent two-site pair blocks and, for odd $L$, one central-site block:
\begin{equation}
\rho(t)=\bigotimes_{j=1}^{M}\rho_{j,L+1-j}(t)\otimes \rho_{\rm c}(t).
\label{eq:rho_factorization}
\end{equation}
This factorization reduces the many-body problem to a finite collection of universal one- and two-site building blocks. Throughout, we take the initial state to be the tilted ferromagnet
\begin{equation}
\ket{\psi_\theta}=\bigotimes_{i=1}^{L}\left(
\cos \left(\frac{\theta}{2}\right)\ket{+1/2}_i+
\sin \left(\frac{\theta}{2}\right)\ket{-1/2}_i
\right),
\label{eq:tilted_initial_state}
\end{equation}
and we use the abbreviations
\begin{equation}
c=\cos\frac{\theta}{2},\qquad
s=\sin\frac{\theta}{2},\qquad
\Delta=\sqrt{1-(\gamma/4)^2},
\label{eq:notation_c_s_delta}
\end{equation}
where the dissipator $L_i = S^z_i$ is the same as in main-text Eq.~(\ref{eq:lindblad}). Since $(S^z)^2 = \mathbb{1}/4$, the dissipator simplifies to $\gamma(S_i^z\rho\, S_i^z - \rho/4)$, so the effective dephasing rate appearing in the analytical formulas below is $\gamma/4$. The parameter $\Delta$ is real, and the dynamics underdamped, for $\gamma < 4$; the overdamped regime $\Delta = i\sqrt{(\gamma/4)^2-1}$ sets in at $\gamma > 4$.

For a single mirror pair, we use the ordered basis
$\left[\ket{+1/2,+1/2},\ket{+1/2,-1/2},
\ket{-1/2,+1/2},\ket{-1/2,-1/2}\right]$.
On this basis, the pair Hamiltonian is
\begin{equation}
H_{\rm pair}=S^+\otimes S^+ + S^-\otimes S^-
=\begin{pmatrix}
0&0&0&1\\
0&0&0&0\\
0&0&0&0\\
1&0&0&0
\end{pmatrix},
\label{eq:pair_hamiltonian}
\end{equation}
while the two dephasing jumps are $S^z\otimes I$ and $I\otimes S^z$. The initial pair state is
\begin{equation}
\ket{\psi_\theta^{(2)}}
=c^2\ket{+1/2,+1/2}+cs\ket{+1/2,-1/2}
+cs\ket{-1/2,+1/2}+s^2\ket{-1/2,-1/2}.
\label{eq:pair_initial_state}
\end{equation}
As $H_{\rm pair}$ couples only $\ket{+1/2,+1/2}$ and $\ket{-1/2,-1/2}$, the exact pair density matrix has the form
\begin{equation}
\rho_{\rm pair}(t)=
\begin{pmatrix}
a(t) & b^*(t) & b^*(t) & u^*(t)\\
b(t) & p & q(t) & d(t)\\
b(t) & q(t) & p & d(t)\\
u(t) & d^*(t) & d^*(t) & \bar a(t)
\end{pmatrix}.
\label{eq:pair_density_matrix}
\end{equation}
The time-independent diagonal element in the single-spin-flip sector and its coherence are
\begin{equation}
p=\frac{\sin^2\theta}{4},\qquad
q(t)=\frac{\sin^2\theta}{4}e^{-\gamma t}.
\label{eq:p_q_definitions}
\end{equation}
The population imbalance between $\ket{+1/2,+1/2}$ and $\ket{-1/2,-1/2}$ is encoded by
\begin{equation}
z(t)=e^{-\gamma t/2}\cos\theta\left[
\cos(2\Delta t)+\frac{\gamma/4}{\Delta}\sin(2\Delta t)
\right],
\label{eq:z_definition}
\end{equation}
so that
\begin{equation}
a(t)=\frac{3+\cos2\theta}{8}+\frac{z(t)}{2},\qquad
\bar a(t)=\frac{3+\cos2\theta}{8}-\frac{z(t)}{2}.
\label{eq:a_abar_definitions}
\end{equation}
The coherence between the two fully polarized states is
\begin{equation}
u(t)=\frac{\sin^2\theta}{4}e^{-\gamma t}
+i\frac{\cos\theta}{2\Delta}e^{-\gamma t/2}\sin(2\Delta t),
\label{eq:u_definition}
\end{equation}
and the coherences involving the one-spin-flip sector are
\begin{eqnarray}
b(t)&=&\frac{\sin\theta}{4}e^{-\gamma t/2-it}\left(e^{2it}+\cos\theta\right),\nonumber\\
d(t)&=&\frac{\sin\theta}{4}e^{-\gamma t/2-it}\left(e^{2it}-\cos\theta\right).
\label{eq:b_d_definitions}
\end{eqnarray}
Equations \eqref{eq:pair_density_matrix}–\eqref{eq:b_d_definitions} are the elementary dynamical building blocks from which all subsystem charged moments are constructed.

Now, tracing out either site of a mirror pair gives the same one-site reduced state,
\begin{equation}
\rho_{\rm red}(t)=
\begin{pmatrix}
\frac{1+z(t)}{2} & \kappa(t)\\
\kappa^*(t) & \frac{1-z(t)}{2}
\end{pmatrix},
\label{eq:one_site_reduced_state}
\end{equation}
with
\begin{equation}
\kappa(t)=\frac{\sin\theta}{2}e^{-\gamma t/2}
\left(\cos t+i\cos\theta\sin t\right).
\label{eq:kappa_definition}
\end{equation}
Its purity is therefore
\begin{equation}
\mathrm{Tr}(\rho_{\rm red}^2)
=\frac{1+z(t)^2}{2}+2|\kappa(t)|^2.
\label{eq:red_purity}
\end{equation}

The central site is not acted on by the Hamiltonian and It undergoes only local dephasing, giving
\begin{equation}
\rho_{\rm c}(t)=
\begin{pmatrix}
\cos^2\frac{\theta}{2} & \frac{1}{2}e^{-\gamma t/2}\sin\theta\\
\frac{1}{2}e^{-\gamma t/2}\sin\theta & \sin^2\frac{\theta}{2}
\end{pmatrix}.
\label{eq:central_density_matrix}
\end{equation}

For a subsystem A, the second Rényi charge asymmetry is obtained from the charged moment
\begin{equation}
Z_A^{(c)}(\alpha,t)=
\mathrm{Tr}\left[\rho_A(t)e^{i\alpha Q_A}\rho_A(t)e^{-i\alpha Q_A}\right],
\label{eq:charge_charged_moment}
\end{equation}
where $Q_A$ is the subsystem charge. We define
\begin{equation}
\Delta S_{A,c}^{(2)}(t)=
-\log\int_{-\pi}^{\pi}\frac{d\alpha}{2\pi}
\frac{Z_A^{(c)}(\alpha,t)}{Z_A^{(c)}(0,t)}.
\label{eq:charge_asymmetry_definition}
\end{equation}
Due to the factorization in Eq.~\eqref{eq:rho_factorization}, $Z_A^{(c)}$ is a product of independent block contributions.

If A contains only one site from a mirror pair, the relevant block is $\rho_{\rm red}(t)$. With $S^z$ as the local charge generator, one obtains
\begin{equation}
g(\alpha)=
\mathrm{Tr}\left[\rho_{\rm red}(t)e^{-i\alpha S^z}
\rho_{\rm red}(t)e^{i\alpha S^z}\right]=\frac{1+z(t)^2}{2}+2|\kappa(t)|^2\cos\alpha,=\mathcal{A}(t) + \mathcal{B}(t)\cos\alpha.
\label{eq:g_single_site_charge}
\end{equation}
Using Eq.~\eqref{eq:kappa_definition}, this may equivalently be written as
\begin{equation}
g(\alpha)=\frac{1+z(t)^2}{2}
+\frac{1}{2}e^{-\gamma t}\sin^2\theta
\left(1-\sin^2\theta\sin^2 t\right)\cos\alpha.
\label{eq:g_single_site_charge_explicit}
\end{equation}

If A contains both sites of a mirror pair, the charge operator on the pair is
\begin{equation}
Q_{\rm pair}^{(c)}=\mathrm{diag}(1,0,0,-1).
\label{eq:charge_pair_operator}
\end{equation}
The corresponding block charged moment is
\begin{equation}
g_2(\alpha)=
\mathrm{Tr}\left[\rho_{\rm pair}(t)e^{i\alpha Q_{\rm pair}^{(c)}}
\rho_{\rm pair}(t)e^{-i\alpha Q_{\rm pair}^{(c)}}\right].
\label{eq:g2_charge_definition}
\end{equation}
In terms of the matrix elements in Eq.~\eqref{eq:pair_density_matrix},
\begin{equation}
g_2(\alpha)=
a(t)^2+\bar a(t)^2+2p^2+2q(t)^2+4\left(|b(t)|^2+|d(t)|^2\right)\cos\alpha
+2|u(t)|^2\cos(2\alpha).
\label{eq:g2_charge_explicit}
\end{equation}
This expression makes explicit the charge-sector structure: the one-unit charge-changing coherences produce $\cos\alpha$, while the two-unit coherence $u(t)$ produces $\cos(2\alpha)$.

For the central site, one can similarly show that
\begin{equation}
g_3(\alpha)=
\mathrm{Tr}\left[\rho_{\rm c}(t)e^{-i\alpha S^z}
\rho_{\rm c}(t)e^{i\alpha S^z}\right]=\frac{1}{4}\left(3+\cos2\theta
+2e^{-\gamma t}\sin^2\theta\cos\alpha\right).
\label{eq:g3_charge}
\end{equation}

Collecting all these facts, let A contain $N_1$ sites whose mirror partners are outside $A$, $N_2$ complete mirror pairs, and $N_c$ central sites, which thus leads to
\begin{equation}
Z_A^{(c)}(\alpha,t)=
g(\alpha)^{N_1}g_2(\alpha)^{N_2}g_3(\alpha)^{N_c}.
\label{eq:charge_factorized_moment}
\end{equation}
Consequently,
\begin{equation}
\Delta S_{A,c}^{(2)}(t)=
-\log\int_{-\pi}^{\pi}\frac{d\alpha}{2\pi}
\frac{g(\alpha)^{N_1}g_2(\alpha)^{N_2}g_3(\alpha)^{N_c}}
{g(0)^{N_1}g_2(0)^{N_2}g_3(0)^{N_c}}.
\label{eq:charge_general_asymmetry}
\end{equation}
Two subsystem geometries used below are obtained as special cases. For a left-half geometry containing three sites from distinct mirror pairs,
\begin{equation}
Z_A^{(c)}(\alpha,t)=g(\alpha)^3.
\label{eq:charge_left_half_example}
\end{equation}
For a bulk geometry containing one complete mirror pair and the central site,
\begin{equation}
Z_A^{(c)}(\alpha,t)=g_2(\alpha)g_3(\alpha).
\label{eq:charge_bulk_example}
\end{equation}

The same construction applies to the dipole asymmetry. We define the subsystem dipole operator by
\begin{equation}
D_A=\sum_{i\in A}(i-i_0)S_i^z,\qquad i_0=\frac{L+1}{2},
\label{eq:dipole_operator}
\end{equation}
and the corresponding charged moment by
\begin{equation}
Z_A^{(d)}(\alpha,t)=
\mathrm{Tr}\left[\rho_A(t)e^{i\alpha D_A}\rho_A(t)e^{-i\alpha D_A}\right].
\label{eq:dipole_charged_moment}
\end{equation}
The second R'enyi dipole asymmetry is
\begin{equation}
\Delta S_{A,d}^{(2)}(t)=
-\log\int_{-\pi}^{\pi}\frac{d\alpha}{2\pi}
\frac{Z_A^{(d)}(\alpha,t)}{Z_A^{(d)}(0,t)}.
\label{eq:dipole_asymmetry_definition}
\end{equation}
Compared with the charge case, the only change is that each off-diagonal matrix element acquires a phase determined by the dipole-charge difference between the two basis states.

For a single site at distance $l$ from the center,
\begin{equation}
g_d(\alpha;l)=
\mathrm{Tr}\left[\rho_{\rm red}(t)e^{i l\alpha S^z}
\rho_{\rm red}(t)e^{-i l\alpha S^z}\right]=\frac{1+z(t)^2}{2}+2|\kappa(t)|^2\cos(l\alpha),=\mathcal{A}(t) + \mathcal{B}(t)\cos (l\alpha).
\label{eq:gd_single_site}
\end{equation}

For a complete mirror pair at distance n from the center, the dipole operator in the pair basis is
\begin{equation}
Q_{\rm pair}^{(d)}=\mathrm{diag}(0,-n,+n,0).
\label{eq:dipole_pair_operator}
\end{equation}
Therefore
\begin{equation}
g_{2d}(\alpha;n)=
\mathrm{Tr}\left[\rho_{\rm pair}(t)e^{i\alpha Q_{\rm pair}^{(d)}}
\rho_{\rm pair}(t)e^{-i\alpha Q_{\rm pair}^{(d)}}\right].
\label{eq:g2d_definition}
\end{equation}
Using Eq.~\eqref{eq:pair_density_matrix}, one finds
\begin{equation}
g_{2d}(\alpha,n)=
a(t)^2+\bar a(t)^2+2p^2+2|u(t)|^2
+4\left(|b(t)|^2+|d(t)|^2\right)\cos(n\alpha)
+2q(t)^2\cos(2n\alpha).
\label{eq:g2d_explicit}
\end{equation}
Here the coherence $q(t)$, which connects $\ket{+1/2,-1/2}$ and $\ket{-1/2,+1/2}$, changes the dipole charge by $2n$, while $b(t)$ and $d(t)$ change it by $n$. The coherence $u(t)$ connects states with equal dipole charge and therefore contributes to the $\alpha$-independent part.

The central site has zero dipole weight. Hence, its dipole charged moment is independent of $\alpha$:
\begin{equation}
g_{3d}(\alpha)=\mathrm{Tr},\rho_{\rm c}(t)^2
=\frac{1}{4}\left(3+\cos2\theta+2e^{-\gamma t}\sin^2\theta\right).
\label{eq:g3d_central}
\end{equation}

Suppose the sites in A whose mirror partners are outside A have distances $l_1,l_2,\ldots$ from the center, and suppose the complete mirror pairs in A have pair distances $n_1,n_2,\ldots$. Then the dipole charged moment factorizes as
\begin{equation}
Z_A^{(d)}(\alpha,t)=
\prod_j g_d(\alpha;l_j)
\prod_k g_{2d}(\alpha;n_k)
g_{3d}(\alpha)^{N_c}.
\label{eq:dipole_factorized_moment}
\end{equation}
The second-R'enyi dipole asymmetry is therefore
\begin{equation}
\Delta S_{A,d}^{(2)}(t)=
-\log\int_{-\pi}^{\pi}\frac{d\alpha}{2\pi}
\frac{
\prod_j g_d(\alpha;l_j)
\prod_k g_{2d}(\alpha;n_k)
g_{3d}(\alpha)^{N_c}}
{
\prod_j g_d(0;l_j)
\prod_k g_{2d}(0;n_k)
g_{3d}(0)^{N_c}}.
\label{eq:dipole_general_asymmetry}
\end{equation}
For the left-half geometry with three single-site contributions,
\begin{equation}
Z_A^{(d)}(\alpha,t)=\prod_{l=1}^{3}g_d(\alpha;l).
\label{eq:dipole_left_half_example}
\end{equation}
For the bulk geometry containing one full pair at distance n=1 and the central site,
\begin{equation}
Z_A^{(d)}(\alpha,t)=g_{2d}(\alpha;1)g_{3d}(\alpha).
\label{eq:dipole_bulk_example}
\end{equation}
\begin{figure}[htb]
\includegraphics[width=0.65\hsize]{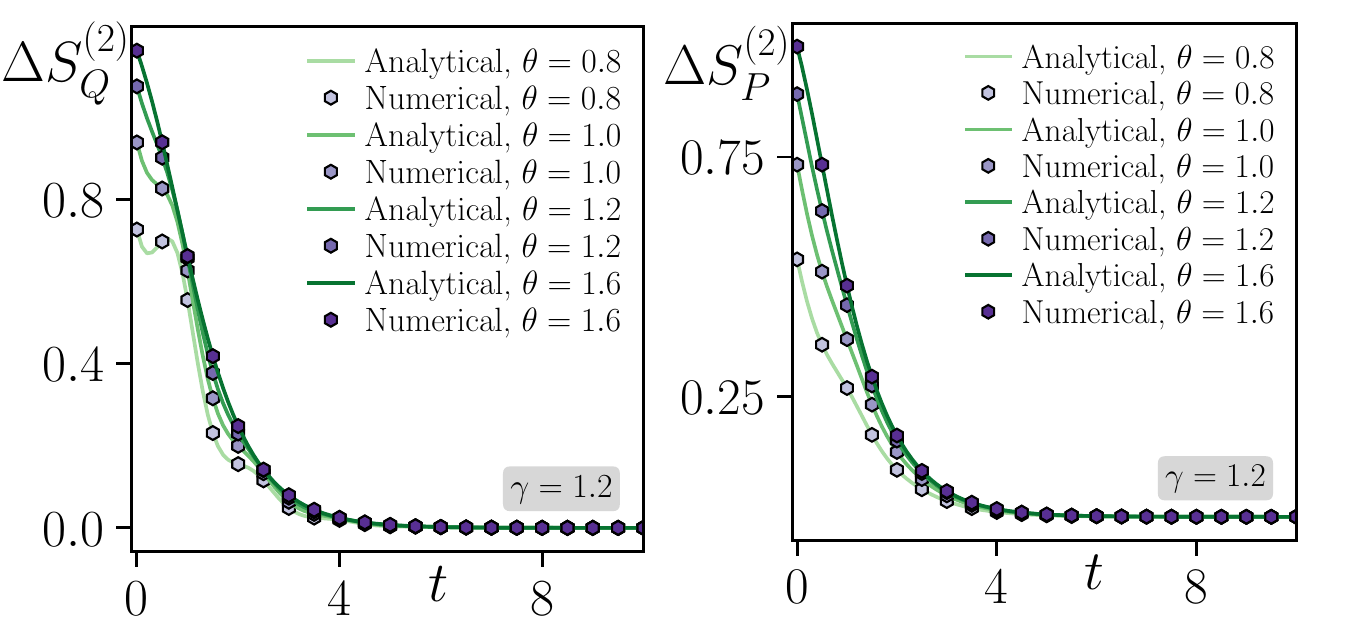}
\caption{For the symmetric pair-flipping model at $\gamma = 1.2$, the R\'enyi-2 asymmetries associated with charge and dipole moment obtained via exact vector simulation for the subsystem comprising $j=-1,0,+1$ show perfect agreement with the analytical results.}
\label{fig:analytical_validation}
\end{figure}
In Fig. \ref{fig:analytical_validation}, we also validate the analytical closed form expressions with vector simulation results for $\gamma=1.2$ and subsystem comprising $j=-1,0,+1$, which show perfect agreement with one another.
\subsection{Crossing time for a general one-sided subsystem}
The closed-form charged moments derived above admit a perturbative analysis of the QME crossing time that extends naturally to a one-sided subsystem $A = \{1, 2, \ldots, L_A\}$ of arbitrary size, including the macroscopic case $L_A = 100$ used in the main text. We show below that the analysis is intrinsically tied to the dissipative regime $\gamma > 0$ and yields a closed-form Mpemba time that is independent of both $L_A$ and the symmetry sector $O \in \{Q, P\}$.

For $A = \{1, 2, \ldots, L_A\}$ with all mirror partners outside $A$, the charged moments factorize over $L_A$ independent single-site building blocks. Since the charge operator $Q_A = \sum_{i \in A} S^z_i$ assigns the same weight to every site, the charge charged moment is the $L_A$-th power of a single factor,
\begin{equation}
Z^{(c)}_A(\alpha, t) \;=\; \bigl[\mathcal{A}(t) + \mathcal{B}(t)\cos\alpha\bigr]^{L_A},
\label{eq:ZcA_LA}
\end{equation}
while the dipole operator $P_A = \sum_{i \in A}\,i\,S^z_i$ assigns a position-dependent weight, so each factor inherits a different phase rate $\ell$,
\begin{equation}
Z^{(d)}_A(\alpha, t) \;=\; \prod_{\ell=1}^{L_A}\bigl[\mathcal{A}(t) + \mathcal{B}(t)\cos(\ell\alpha)\bigr],
\label{eq:ZdA_LA}
\end{equation}
with $\mathcal{A}(t)$ and $\mathcal{B}(t)$ the universal single-site amplitudes of Eq.~\eqref{eq:gd_single_site}. The denominator in both sectors is $Z^{(O)}_A(0, t) = [\mathcal{A} + \mathcal{B}]^{L_A}$.

We now define the dissipative small parameter $\eta(t) \;\equiv\; \frac{\mathcal{B}(t)}{\mathcal{A}(t)}$.
For $\gamma > 0$ one has $\mathcal{B}(t) \propto e^{-\gamma t}$ and $\mathcal{A}(t) \to 1/2$ on the time scale $\gamma^{-1}$, so $\eta(t) \to 0$ exponentially. This is the small parameter that controls the expansion at late times. At $\gamma = 0$ the envelope is absent, $\eta(t)$ remains $O(1)$, and the perturbative scheme below does not apply.

Factoring $\mathcal{A}$ from each building block, the asymmetry ratio takes the form
\begin{equation}
R^{(O)}(t) = \int_{-\pi}^{\pi}\!\frac{d\alpha}{2\pi}\,\frac{Z^{(O)}_A(\alpha, t)}{Z^{(O)}_A(0, t)} = \frac{1}{(1+\eta)^{L_A}}\int_{-\pi}^{\pi}\!\frac{d\alpha}{2\pi}\,\prod_{\ell \in \mathcal{L}_O}\bigl[1 + \eta\cos(\ell\alpha)\bigr]\nonumber\\,
\label{eq:ROt}
\end{equation}
with $\mathcal{L}_Q = \{\underbrace{1, 1, \ldots, 1}_{L_A}\}$ for charge and $\mathcal{L}_P = \{1, 2, \ldots, L_A\}$ for dipole. Expanding the product and integrating term-by-term, only subsets $S \subseteq \mathcal{L}_O$ whose indices admit a signed combination $\sum_{\ell \in S}\sigma_\ell\,\ell = 0$ (with $\sigma_\ell \in \{\pm1\}$) survive the $\alpha$-integration. The lowest non-trivial contribution comes from $|S| = 2$, but the two sectors differ in how it arises:
\begin{itemize}
\item \emph{Charge sector}: $\mathcal{L}_Q$ contains $L_A$ identical indices, so each pair $\{1, 1\}$ with opposite signs cancels. The $\binom{L_A}{2}$ such pairs each contribute $\eta^2/2$,
\begin{equation}
\int_{-\pi}^{\pi}\!\frac{d\alpha}{2\pi}\,[1+\eta\cos\alpha]^{L_A} \;=\; 1 + \binom{L_A}{2}\,\frac{\eta^2}{2} + O(\eta^4).
\end{equation}
\item \emph{Dipole sector}: $\mathcal{L}_P$ contains distinct indices, so no $|S|=2$ subset cancels. The first non-trivial contribution requires $|S| \geq 3$ subsets with vanishing signed sum (e.g., $\{1, 2, 3\}$ with signs $\{-,-,+\}$),
\begin{equation}
\int_{-\pi}^{\pi}\!\frac{d\alpha}{2\pi}\,\prod_{\ell=1}^{L_A}[1+\eta\cos(\ell\alpha)] \;=\; 1 + O(\eta^3).
\end{equation}
\end{itemize}
Expanding the denominator $(1+\eta)^{-L_A} = 1 - L_A\eta + \binom{L_A+1}{2}\eta^2 + O(\eta^3)$ and combining,
\begin{eqnarray}
R^{(c)}(t) &=& 1 - L_A\eta + \frac{3L_A(L_A-1)}{4}\eta^2 + O(\eta^3), \nonumber\\ R^{(d)}(t)& = & 1 - L_A\eta + \frac{L_A(L_A-1)}{2}\eta^2 + O(\eta^3).
\end{eqnarray}
Taking the negative logarithm yields the leading-order R\'enyi-2 asymmetry
\begin{equation}
\Delta S^{(2)}_{A, O}(\theta, t) \;\simeq\; L_A\,\eta(t) \;+\; O\!\bigl(L_A^2\,\eta^2\bigr), \qquad O \in \{Q, P\},\;
\label{eq:DS_LA_leading}
\end{equation}
identical at leading order in both sectors. The asymmetry scales linearly in $L_A$, a direct consequence of the additive log structure of the factorized charged moment. The $O(L_A^2 \eta^2)$ correction sets the regime of validity: the leading-order formula is accurate when $L_A\,\eta(t) \lesssim 1$. For the parameters $\gamma = 1.6$ and $L_A = 100$ used in Fig.~\ref{fig:pairflipping}, this requires $t \gtrsim 3$, where $\eta(t) \lesssim 0.005$ ensures $L_A\eta \lesssim 0.5$.

The Mpemba crossing $\Delta S^{(2)}_{A, O}(\theta_1, t_M) = \Delta S^{(2)}_{A, O}(\theta_2, t_M)$ reduces, at leading order, to $L_A\,\eta(\theta_1, t_M) = L_A\,\eta(\theta_2, t_M)$, i.e.,
\begin{equation}
\frac{\mathcal{B}(\theta_1, t_M)}{\mathcal{A}(\theta_1, t_M)} \;=\; \frac{\mathcal{B}(\theta_2, t_M)}{\mathcal{A}(\theta_2, t_M)}.
\label{eq:crossing_LA}
\end{equation}
\emph{The factor $L_A$ cancels exactly on both sides.} The leading-order Mpemba time is therefore independent of $L_A$ and of the symmetry sector. To extract its explicit form, we use the dissipative-regime approximation $e^{-\gamma t_M} \ll 1$ to set $\mathcal{A}(\theta, t_M) \simeq 1/2$ uniformly in $\theta$, reducing Eq.~\eqref{eq:crossing_LA} to a condition on $\mathcal{B}$ alone:
\begin{equation}
\sin^2\theta_1\bigl(1 - \sin^2\theta_1 \sin^2 t_M\bigr) \;=\; \sin^2\theta_2\bigl(1 - \sin^2\theta_2 \sin^2 t_M\bigr).
\label{eq:Bcondition}
\end{equation}
Setting $s_i \equiv \sin^2\theta_i$ and $\xi \equiv \sin^2 t_M$, Eq.~\eqref{eq:Bcondition} reads $s_1 - s_1^2 \xi = s_2 - s_2^2 \xi$, equivalently $(s_1 - s_2) = (s_1 + s_2)(s_1 - s_2)\xi$. For $\theta_1 \neq \theta_2$ we divide by $(s_1 - s_2)$ to obtain $\xi = 1/(s_1 + s_2)$, hence
\begin{equation}
t_M \;=\; \arcsin\!\left(\frac{1}{\sqrt{\sin^2\theta_1 + \sin^2\theta_2}}\right),
\label{eq:tM_LA}
\end{equation}
valid for all $L_A$ and for both charge and dipole sectors. A real solution requires $\sin^2\theta_1 + \sin^2\theta_2 \geq 1$, providing a lower bound on the breaking strengths needed for the curves to intersect at all.
Table~\ref{tab:tM_LA} compares the closed-form prediction Eq.~\eqref{eq:tM_LA} with the first crossing of the full closed-form asymmetry, obtained by numerical $\alpha$-integration of Eqs.~\eqref{eq:ZcA_LA}-\eqref{eq:ZdA_LA}, for $L_A \in \{2, 5, 10, 50, 100\}$. The predicted value matches the exact first crossing to better than $2\%$ across all $L_A$ and both sectors, validating the $L_A$-independence claim at macroscopic subsystem size. 

\begin{table}[h]
\centering
\caption{Mpemba time at $\gamma = 1.6$, $(\theta_1, \theta_2) = (1.6, 1.2)$: closed-form prediction Eq.~\eqref{eq:tM_LA} versus the first crossing of the exact charge- and dipole-sector R\'enyi-2 asymmetries, obtained by numerical $\alpha$-integration. The $L_A$-independence of $t_M$ is verified to four decimal places.}
\label{tab:tM_LA}
\begin{tabular}{c|ccccc|c}
\hline\hline
$L_A$ & 2 & 5 & 10 & 50 & 100 & formula \\
\hline
$t_M$ (charge) & 0.835 & 0.835 & 0.835 & 0.835 & 0.835 & \multirow{2}{*}{0.821} \\
$t_M$ (dipole) & 0.835 & 0.835 & 0.835 & 0.835 & 0.835 & \\
\hline\hline
\end{tabular}
\end{table}
\subsection{Multiple crossings and the role of dissipation}
The asymmetry curves at $\gamma = 1.6$ exhibit \emph{multiple} crossings at $t_M$ and at the periodic images of Eq.~\eqref{eq:tM_LA}, due to the residual oscillatory factor $1 - \sin^2\theta \sin^2 t$ in $\mathcal{B}(t)$. The Mpemba time corresponds to the \emph{first} such crossing; the subsequent re-crossings are sub-dominant relative to the exponential envelope $e^{-\gamma t}$ and do not affect the conclusion that the more strongly tilted state restores symmetry faster on average. The role of dissipation in the QME of this model is therefore twofold. First, it underwrites the perturbative scheme: at $\gamma = 0$ the building block $\mathcal{B}(t)$ oscillates without an exponential envelope, $\eta(t)$ stays $O(1)$ throughout, and the leading-order expansion of Eq.~\eqref{eq:DS_LA_leading} cannot be invoked. Second, even where Eq.~\eqref{eq:Bcondition} is formally solved at $\gamma = 0$, the asymmetry curves cross repeatedly at the times prescribed by Eq.~\eqref{eq:tM_LA} and its periodic images, lacking any monotonic decay over time. For $\gamma > 0$ the first crossing occurs at $t_M$, and the envelope $e^{-\gamma t}$ damps the subsequent oscillations of $\mathcal{B}(t)$, so the strongly-tilted curve remains parametrically below the weakly-tilted one for $t > t_M$.

\section{Von-Neumann asymmetry study for the symmetric Pair-flipping model}
\label{app:vonNeumann_pair_flipp}

\begin{figure*}
\includegraphics[width=0.96\textwidth]{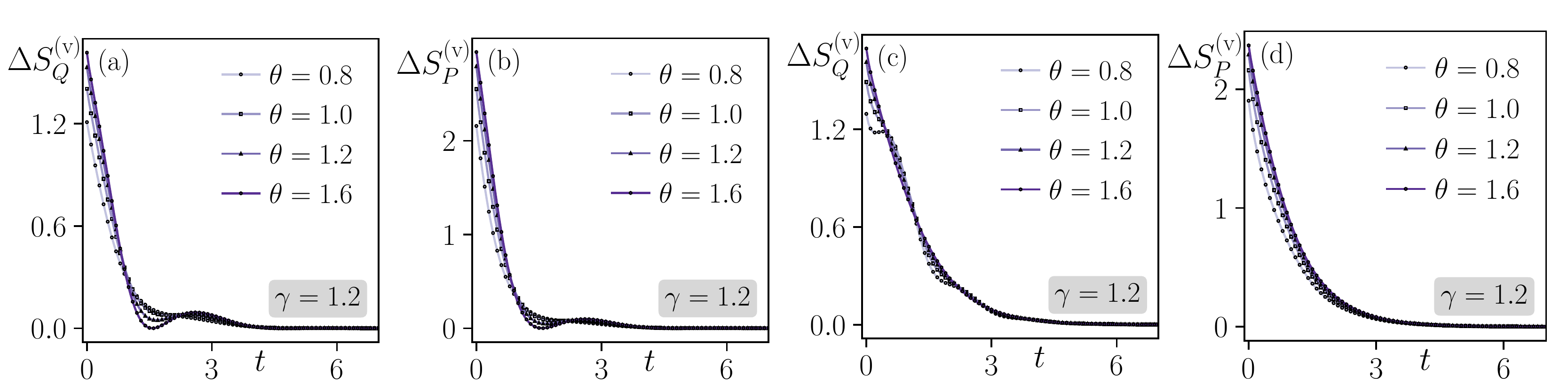}
\caption{Von-Neumann asymmetries with respect to charge and dipole symmetries in the symmetric pair-flipping model.
(a,b) Asymmetries for the subsystem comprising sites $j=1,2,3,4,5,6$, with the central site taken to be $j=0$, for $\gamma=1.2$.
(c,d) Same as panels (a,b), but now subsystems including the central sites, i.e, $j=-3,-2,-1,0,1,2,3 $ for $\gamma=1.2$.
In the dissipative regime, both asymmetries decay, indicating restoration of the corresponding symmetries, including
multiple crossings for (a-c) and no crossings for (d).}
\label{von_neumann_asymmetry}
\end{figure*}

The analytical treatment of the symmetric pair-flipping model in Appendix~\ref{app:pair_factorized_dynamics} was carried out at R\'enyi index $n=2$, for which the charged moment $Z^{(O)}_A(\alpha,t)$ factorizes algebraically over the mirror-pair building blocks of Eqs.~\eqref{eq:pair_density_matrix}--\eqref{eq:b_d_definitions}. The QME, however, is properly defined in terms of the von Neumann asymmetry $\Delta S^{(v)}_{A,O}\equiv \lim_{n\to 1}\Delta S^{(n)}_{A,O}$, since only the $n\to 1$ limit corresponds to the relative entropy between $\rho_A$ and its symmetrisation, as discussed in Sec.~III. In this appendix, we verify that the von Neumann diagnostic reproduces the same qualitative features already established at $n=2$, supporting the use of the experimentally more accessible R\'enyi-$2$ asymmetries throughout the main text.

We compute $\Delta S^{(v)}_{A,O}(t)$ by exact vector simulation of the Lindblad dynamics generated by the Hamiltonian of Eq.~\eqref{eq:mirror_hamiltonian} and the local dephasing jumps $L_x=S^z_x$, exploiting the factorization of the time-evolved density matrix into mirror-pair blocks [Eq.~\eqref{eq:rho_factorization}] to reduce the cost to that of independent two-site sectors. Figure~\ref{von_neumann_asymmetry} shows the resulting dynamics at $\gamma=1.2$ for tilted ferromagnetic initial states with $\theta\in\{0.8,1.0,1.2,1.6\}$, in the same two subsystem geometries considered in the R\'enyi-2 analysis.

The von Neumann data agree qualitatively with the R\'enyi-2 phenomenology of Fig.~\ref{fig:pairflipping}(c-d,g-h). In all four panels, both $\Delta S^{(v)}_{A,Q}(t)$ and $\Delta S^{(v)}_{A,P}(t)$ decay toward zero on a timescale set by the dephasing rate, confirming that the local dephasing dresses the coherent pair-flip dynamics with an exponential envelope $e^{-\gamma t}$ that drives the relevant amplitudes [Eqs.~\eqref{eq:z_definition}--\eqref{eq:kappa_definition}] toward their symmetric values and hence restores both symmetries. For the subsystem $A=\{1,\dots,6\}$, which excludes the central site and samples six independent mirror-pair distances, panels (a-b) display clear multiple Mpemba-like crossings in both sectors. For the symmetric subsystem $A=\{-3,\dots,3\}$, which contains the central site together with three complete mirror pairs, the charge sector in panel (c) still exhibits multiple crossings, whereas the dipole sector in panel (d) preserves the initial $\theta$-ordering throughout the relaxation. These features are completely consistent with the Renyi-2 case.

Taken together, the data of Fig.~\ref{von_neumann_asymmetry} establish that the qualitative features of the higher-order QME in the symmetric pair-flipping model, restoration of both symmetries and multiple crossings, are independent of the R\'enyi index used as a diagnostic, and persist at the level of the von Neumann asymmetry that conceptually underpins the definition of the QME.

\bibliography{ref_mpemba}

\end{document}